\documentclass{pasj00}
\draft

\begin{document}
\SetRunningHead{Kaneda et al.}{Dust in Hot Plasma of Elliptical Galaxies}
\Received{//}
\Accepted{//}

\title{Dust in Hot Plasma of Nearby Dusty Elliptical Galaxies Observed with the Spitzer Space Telescope}

\author{%
   Hidehiro \textsc{Kaneda},\altaffilmark{1}
   Takashi \textsc{Onaka},\altaffilmark{2}
   Tetsu \textsc{Kitayama},\altaffilmark{3}\\
   Yoko \textsc{Okada},\altaffilmark{1}
   and
   Itsuki \textsc{Sakon}\altaffilmark{2}}
 \altaffiltext{1}{Institute of Space and Astronautical Science, \\
Japan Aerospace Exploration Agency, Sagamihara, Kanagawa 229-8510}
 \email{kaneda@ir.isas.jaxa.jp}
 \altaffiltext{2}{Department of Astronomy, Graduate School of Science, University of Tokyo, \\
Bunkyo-ku, Tokyo 113-0003}
 \altaffiltext{3}{Department of Physics, Toho University, \\
Funabashi, Chiba 274-8510}


%

\KeyWords{infrared: ISM --- ISM: lines and bands --- ISM: dust, extinction --- 
--- galaxies: elliptical and lenticular, cD --- galaxies: ISM} 

\maketitle

\begin{abstract}
We report on mid- and far-IR {\it Spitzer} observations of 7 nearby dusty elliptical galaxies by using the Multiband Imaging Photometer (MIPS) and Infrared Spectrograph (IRS). Our sample galaxies are known to contain an excessive amount of interstellar dust against sputtering destruction in hot plasma filling the interstellar space of elliptical galaxies. In order to study the origin and the properties of the excess dust in the hot plasma, we selected galaxies with a wide range of X-ray luminosities but similar optical luminosities for our {\it Spitzer} Guest Observers (GO1) program. The 7 galaxies are detected at the MIPS 24 $\mu$m, 70 $\mu$m, and 160 $\mu$m bands; the far- to mid-IR flux ratios of relatively X-ray-bright elliptical galaxies are lower than those of X-ray-faint galaxies. From the IRS spectra, polycyclic aromatic hydrocarbon (PAH) emission features are detected significantly from 5 of the 7 galaxies; the emission intensities are weaker as the X-ray luminosity of the galaxy is larger. We have found a correlation between the far- to mid-IR flux ratio and the equivalent width of the PAH emission feature. We have obtained apparent spatial correspondence between mid-IR and X-ray distributions in the outer regions for the three X-ray-brightest galaxies in our sample. Possible interpretations for our observational results are discussed. 
\end{abstract}
\section{Introduction}
Ordinary elliptical galaxies are deficient in dust as compared with typical spiral galaxies and thus necessarily faint mid- and far-IR emitters. Therefore intensive studies of dust in elliptical galaxies have long been challenging. Profound knowledge of dust physics in elliptical galaxies is, however, extremely valuable because the galaxies provide the dust with a unique environment, i.e., old stellar radiation fields without active star formation and interstellar media (ISM) mostly dominated by hot ionized gas. {\it IRAS} 60 $\mu$m and 100$\mu$m observations detected about half of E and E/S0 galaxies in the RSA catalog \citep{Kna89}, indicating that the presence of cold dust (T=20$\sim$40 K) in elliptical galaxies was common at some level. However, most detections were near the threshold of the instrument, and only about 12 \% of normal elliptical galaxies in the catalog were detected above the 98 \% confidence level \citep{Bre98}.

The most plausible source for the interstellar dust in elliptical galaxies is accumulation of dust ejected by mass-losing giant stars. In elliptical galaxies containing a large volume of hot plasma, however, such interstellar dust is expected to be easily destroyed through sputtering by ambient plasma ions (Draine \& Salpeter 1979; Dwek \& Arendt 1992; Tielens et al. 1994). Hence, there should be some physical constraints on the maximum dust mass for a given blue luminosity within a galaxy, which is determined by the balance between the destruction and the replenishment. Several elliptical galaxies detected with {\it IRAS}, however, are found surprisingly to contain large dust masses which are well beyond such limits \citep{Gou95}. {\it ISO} detected 16 elliptical galaxies of the observed 39 galaxies in 60$-$200 $\mu$m, showing that the dust masses derived from {\it ISO} are even ten times greater on the average than those estimated from {\it IRAS} \citep{Tem04}. How can such a huge amount of the interstellar dust grains survive in these galaxies? {\it ISO} observations have also shown that, in addition to the cold dust, a number of elliptical galaxies contain a measurable amount of warm dust \citep{Fer02}. Recent spectroscopic studies with {\it Spitzer} have revealed even the presence of polycyclic aromatic hydrocarbons (PAHs) emission features for several dusty elliptical galaxies (Kaneda et al. 2005; J. D. Bregman et al. 2006).  Is the interstellar dust really interacting with the plasma? What are the origins of the excess dust and the PAHs?

In this paper, we report on mid- and far-IR observations of 7 nearby dusty elliptical galaxies by using the Multiband Imaging Photometer (MIPS; Rieke et al. 2004) and Infrared Spectrograph (IRS; Houck et al. 2004) on board {\it Spitzer} \citep{Wer04}, which are the whole of our Guest Observers (GO1) program (ID: 3619; PI: H. Kaneda). These 7 galaxies are among the {\it IRAS} dusty elliptical galaxies described in \citet{Gou95}; dust masses derived from the {\it IRAS} flux densities exceed by more than one order of magnitude the threshold where dust is replenished by stellar mass loss at the rate given by Faber \& Gallagher (1976) and destroyed for the destruction time scale of $10^7$ yr. This time scale corresponds to a typical value for the sputtering destruction in elliptical galaxies known to contain X-ray-emitting gas (see, e.g., Canizares et al. 1987). The objective of our observations is to study the origins and the properties of the excess dust and the PAHs in the hot plasma environment. The properties of the observed galaxies are summarized in Table 1. Among them, NGC~1407 is the central galaxy of the cluster of galaxies, Eridanus A, which is one of the darkest known clusters \citep{Gou93}, while NGC~4696 is the central galaxy in the Centaurus Cluster; both galaxies are well studied in X-rays (see, e.g., Zhang \& Xu 2004; Fabian et al. 2005). It is confirmed from the table that our sample galaxies are similar in morphological types, sizes, and optical luminosities but quite different in X-ray luminosities, and therefore ideal for investigating effects of the hot plasma on the dust. 

\section{Observations and Data Analyses}
The MIPS and IRS observations of the 7 elliptical galaxies, NGC~~1395, NGC~~1407, NGC~~2974, NGC~~3962, IC~3370, NGC~~4589, and NGC~~4696 were carried out in our GO1 program from 2004 Oct to 2005 Sep, except for the MIPS observations of NGC~4696. The summary of the observation log is listed in Table 2. For each galaxy, we obtained 24 $\mu$m, 70 $\mu$m, and 160 $\mu$m band images using the MIPS in the large-source photometry mode and 5$-$14 $\mu$m spectra using the IRS/Short-Low (SL) module in the standard staring mode. Each target was observed as a fixed single for the MIPS 24 $\mu$m and 160 $\mu$m observations and as two fixed cluster offsets for the MIPS 70 $\mu$m observation with an ``offsets ($\pm 80''$) only'' option  and for the IRS observations. 
The MIPS observations of NGC~4696 were performed within the framework of another GO1 program by using the small-source photometry mode with rather short exposure time; we used the archives data.

The mosaic images were made from the MIPS Basic Calibrated Data (BCD; pipeline version S12; Gordon et al. 2005) by using the MOPEX (version 093005) software package \citep{Mak05}.
For the MIPS 24 $\mu$m band, prior to mosaicking, we self-calibrated the BCD images to correct persistent effects including long-term latents and gradients, as recommended in the MIPS Data Handbook; in the large-source photometry mode, a half of BCD were taken from the observations of adjacent background skies, and thus utilized for flat corrections after removing point sources and coadding all the background BCD. For NGC~4696 taken in the small-source photometry mode, we removed the gradients by linearly fitting the data in the outer parts ($>$\timeform{1'.5} from the center) of the galaxy after mosaicking. 
As for the MIPS 70 $\mu$m band, we applied column filtering which is one of the IDL routines available on the {\it Spitzer} Science Center (SSC) website in order to remove streaking due to residual slow response variations. 

The spectra were extracted and calibrated from the IRS BCD (pipeline version: S13) by using the SMART (version 5.5.6) software package \citep{Hig04}. We first operated subtraction in two dimensions of background from signal data for each sub-slit and nod position; a blank sky was observed with a spatial separation of about $2'$ from the center of each galaxy. The first and second orders matched fairly well and required no further base-line adjustment. The ends of each order where the noise increases significantly were manually clipped.

The mosaic images thus obtained for the 7 elliptical galaxies are shown in Figure 1, where the MIPS 24 $\mu$m and 70 $\mu$m maps are smoothed to $\sim 15''$ scales similar to the MIPS 160 $\mu$m maps for comparative purposes. The image size is about \timeform{3'.5}$\times$\timeform{3'.5} for each. The boundaries of the MIPS 160 $\mu$m observations (and also the MIPS 70 $\mu$m observation for NGC~4696) are shown together. The background levels were estimated from nearby blank skies and subtracted from the images.
The IRS/SL spectra are shown in Figure 2, where the data are averaged into the resolution of 0.06 $\mu$m in wavelength. For IC~3370, the observation missed the center of the galaxy by $10''$, which caused significant reduction in the signal. For clarity of presentation, each spectrum is shifted by applying an appropriate gain factor as labeled in the figure. 

\section{Results}
We have obtained the mid- and far-IR images of the 7 elliptical galaxies, most of which are spatially resolved. We have derived the flux densities of the galaxies in the MIPS 24 $\mu$m, 70 $\mu$m, and 160 $\mu$m bands (Table 3). These values are obtained by integrating surface brightness within a diameter of $1.3D_{n}$ from the center of each galaxy, where $D_{n}$ is the optical angular diameter given in Faber et al. (1989) and listed in Table 1. The aperture size is shown by dashed circles in Figure 1. The flux uncertainties are estimated from error images provided with the BCD by adding values within the same integration area as the flux densities are obtained; they do not include either systematic effects associated with the detectors or absolute uncertainties of about 20 \% for MIPS 70 $\mu$m and 160 $\mu$m. The above aperture size is selected because it gives a relatively good singal-to-noise ratio for each galaxy, covering dominant central regions of dust emission and avoiding contaminations from neighboring sources (see, e.g., NGC~1395 and NGC~2974). Aperture corrections are performed for MIPS 70 $\mu$m and 160 $\mu$m by using correction tables available on the SSC website. The MIPS 24 $\mu$m fluxes provide us brand-new information, since most of the elliptical galaxies are given only upper limits in mid-IR before {\it Spitzer}. The values show fairly good agreement with {\it IRAS} and/or {\it ISO} observations except for the MIPS 70 $\mu$m and 160 $\mu$m fluxes of NGC~1407. As seen in Figure 1, there are significant fluctuations in far-IR sky brightness around NGC~1407, which have no clear signs of association with the galaxy; {\it IRAS} would suffer confusion of them due to the low angular resolution. As a result, we have found that cold dust emission from NGC~1407 is much fainter than that measured in the {\it IRAS} observations, at least in the inner parts of the galaxy.

The spectral energy distributions (SEDs) constructed from the flux densities in Table 3 are presented in Figure 3. It is found from the SEDs that the variations of the MIPS 24 $\mu$m flux are noticeably smaller than those of the MIPS 70 $\mu$m and 160 $\mu$m fluxes, which may suggest that dominant carriers for the mid- and far-IR bands are not of the same origin; interstellar dust emission dominates the far-IR bands, whereas circumstellar dust emission (i.e., silicate emission from evolved stars; see, e.g., Bressan et al. 2006; Athey et al. 2002) may significantly contribute to the 24 $\mu$m fluxes; this can reduce the MIPS 24 $\mu$m flux variations because our sample has similar optical fluxes (Table 1) and thus similar stellar contributions to the mid-IR band are expected. Recently, several authors (Athey et al. 2002; Xilouris et al. 2004; Temi et al. 2005) show that mid-IR (6$-$15 $\mu$m) emission in elliptical galaxies follows the de Vaucouleurs $R^{1/4}$ profile of optical starlight. In Figure 4, we plot the MIPS 24 $\mu$m surface brightness profiles in the inner parts of our sample galaxies as a function of the semimajor axis normalized by $R_e$ to the 1/4 power, where $R_e$ is the effective radius; we find that the profiles of most galaxies are in good agreement with a de Vaucouleurs profile that is a straight line with a slope of $-3.25$ \citep{Vau48} in this plot, indicating that the warm dust is mostly of circumstellar origin.  

We have calculated dust masses by using the equation (2) in Goudfrooij \& de Jong (1995), where the grain emissivity factor is given by Hildebrand (1983) and the average grain radius of 0.1 $\mu$m and the specific dust mass density of 3 g cm$^{-3}$ are used. Dust temperatures are set to be equal to the color temperatures determined from the ratios of the MIPS 160 $\mu$m to the MIPS 70 $\mu$m flux under the assumption that the far-IR emission of elliptical galaxies originates from dust with an emissivity law $\propto$ $\lambda^{-1}$. The results are listed in Table 4. Dust temperatures thus derived are consistent with those estimated in \citet{Gou95}. Some galaxies show dust masses significanly smaller than those in \citet{Gou95}, which may be caused by {\it IRAS} source confusion and/or the presence of dust extended beyond the photometry apertures. We should also note that the MIPS is insensitive to colder dust emitting predominantly at submillimeter wavelengths (e.g., Leeuw et al. 2004). Hence the dust masses in Table 4 are considered to be lower limits to the real dust mass contained in the whole galaxy, which ensures the presence of dust with excessive masses for our sample galaxies as described in \citet{Gou95}.

We show the IRS spectra of our sample in Figure 2, where the spectra of NGC~2974, NGC~3962, NGC~4589, and NGC~4696 have already been presented in \citet{Kan05}. We have detected the PAH emission features from 5 out of the 7 elliptical galaxies in our sample. They reveal unusual polycyclic aromatic hydrocarbon (PAH) emission features; the usually strongest emission features at 7.7 $\mu$m (see, e.g., Onaka et al. 1996) are faint in contrast to prominent emission features at 11.3 $\mu$m. As pointed out by J. D. Bregman et al. (2006), a broad 7.7 $\mu$m emission feature may appear to be relatively weak due to the presence of a stellar silicate feature and continuum underlying shortward of 10 $\mu$m. However, the differences from usual PAH emission features observed for our Galaxy, normal, and starburst galaxies seem to be too large to be explained by only considering such effects; the relative integrated intensity of the 11.3 $\mu$m to the 7.7 $\mu$m feature is usually $0.2-0.5$ (Sakon et al. 2004; Brandl et al. 2004; Lu et al. 2003), but is as large as $2-4$ for the elliptical galaxies detected in our sample \citep{Kan05}. We have estimated the intensities of the 11.3 $\mu$m emission features by single Gaussian fits to the spectra within SMART; the results are listed in Table 5.

\subsection{Central emission of dust and PAHs in hot plasma}
We plot the equivalent width of the 11.3 $\mu$m PAH emission feature and the far- to mid-IR flux ratio against the X-ray luminosity of the galaxy in Figures 5a and 5b, respectively. The flux ratio is derived by dividing the average of the MIPS 70 $\mu$m and 160 $\mu$m fluxes by the MIPS 24 $\mu$m flux in Table 3. Since the X-ray luminosity of IC~3370 is not given in any paper, we estimate its upper limit on the basis of the $\log L_{\rm B}$ vs. $\log L_{\rm X}$ scatter plot in O'Sullivan et al. (2001). NGC~4696, the central galaxy of the Centaurus cluster, has a much higher X-ray luminosity than the others. Both parameters seem to have anti-correlations with the X-ray luminosities except that only NGC~4696 in Figure 5b deviates from such trend (linear-correlation coefficient $R=-0.67$ for Fig.5a and $-0.48$ for Fig.5b). The interpretation on the former relation (Fig. 5a) may be straightforward. Since the lifetime of very small particles with sizes of about 1 nm is very short ($10^5\sim 10^6$ yr; Draine \& Salpeter 1979) in hot plasma with typical gas densities of $10^{-2}\sim 10^{-3}$ cm$^{-3}$, PAHs are expected to be easily destroyed, which necessarily more or less produces such anti-correlation. Considering the fact that normal quiescent elliptical galaxies do not exhibit significant PAH emission features (J. N. Bregman et al. 2006), the significant detection of PAHs from 5 out of the 7 elliptical galaxies is rather surprising; the presence of PAHs suggests star formation in the recent past that has been triggered by a merger event (J. D. Bregman et al. 2006). The unusually strong emission feature at 11.3 $\mu$m as compared with the emission features at $<$ 10 $\mu$m may be explained by the dominance of large-size PAHs over small ones (Schutte et al. 1993), which is consistent with the picture of PAHs currently being destroyed in hot plasma because larger PAHs are expected to survive for a longer time against sputtering destruction. Alternatively, the prominent emission feature at 11.3 $\mu$m can be explained by the dominance of neutral PAHs \citep{Job94}.

Since the central mid-IR component is probably dominated by circumstellar dust emission from evolved stars as discussed above, a plausible way to explain the latter relation (Fig. 5b) is to postulate that the circumstellar dust is somewhat unaffected by the hot plasma, while the cold dust responsible for the far-IR emission is under sputtering destruction. We plot the ratio of the MIPS 160 $\mu$m to the MIPS 70 $\mu$m flux against the far- to mid-IR flux ratio in Figure 5c, where we find that the far-IR emission becomes colder as the far-IR flux is lower ($R=-0.81$). This can be explained by the shorter lifetime of smaller grains under sputtering destruction, thus proving that the dust is really interacting with the hot plasma. Although the temperature may also be increased through collisions with plasma electrons, which would cause the opposite trend in Figure 5c, the dominant source of grain heating in the inner parts of elliptical galaxies is absorption of starlight and can hardly be collisions with electrons \citep{Tsa95}.

In Figure 5d, we have found a correlation between the intensities of the PAHs and the cold dust ($R=0.93$). It is true that these two can be correlated via sputtering destruction by the hot plasma, but they are coupled with each other more closely than expected from their dependencies on the X-ray luminosity, suggesting that there is another underlying physics for this strong correlation. A naive interpretation is that the excess cold dust is of the same origin as the observed PAHs; two seperate processes of the replenishment and the destruction control the current amount of the PAHs and the cold dust in the ISM, in which the former process seems to be relatively important to cause such strong correlation. In light of this interpretation, the observed amount of PAHs in NGC4696 seems to be rather small for its relatively large content of cold dust, which may indicate a harsher environment for very small particles in the center of the X-ray-bright cluster of galaxies. As a candidate of their common origin, currently there are two popular concepts for the evolution of the ISM of luminous elliptical galaxies; one is the cooling flow scenario (Fabian 1994), in which, if the heating sources are not strong enough to balance the cooling, the hot gas flows inward where it converts into cooler gas. The other is the evaporation flow scenario (Sparks, Macchetto \& Golombek 1989; de Jong et al. 1990), in which clouds of cold gas and dust have accreted during post-collapse galaxy interactions. Any matter that condenses out of a cooling flow is, however, likely to be devoid of dust grains \citep{Gou98}. From its connection with PAHs that are thought to be a remnant of past star formation triggered by a recent merger, the excess dust is also likely to be brought in by the same merger event rather than internally produced. 
Hence our observational result favors the latter concept for the origin of the excess dust.

\subsection{Extended emission of dust in hot plasma}
Some galaxies show IR emission extended beyond the photometry apertures, which does not contribute to the flux densities in Table 3. As a whole, the MIPS 70 $\mu$m and 160 $\mu$m emission seems to have spatial distributions similar to that of the MIPS 24 $\mu$m emission in the outer regions. However, the accuracies in the surface brightness of such faint far-IR emission extended around the peak are quite uncertain since the MIPS far-IR detectors are reported to suffer severe slow response that is somewhat inherent in far-IR photoconductors. Therefore, we concentrate on more reliable MIPS 24 $\mu$m images below.  

We take the three X-ray-brightest galaxies in our sample, NGC~1395, NGC~1407, and NGC~4696, for which X-ray images can be retrieved from the {\it Chandra} Advanced CCD Imaging Spectrometer (ACIS) archives data.
We superpose the MIPS 24 $\mu$m contour maps on the X-ray images (Fig. 6). In order to obtain the spatial distribution of the hot plasma alone, X-ray point sources that are thought to be low-mass X-ray binaries have been removed from the images by using the {\it Chandra} Interactive Analysis of Observations (CIAO) software. The X-ray images are then smoothed and plotted in logarithmic scale for comparative purposes. In these images, we have found that there are apparent spatial correspondence between the mid-IR and X-ray distributions. More quantitatively, in Figure 7, we plot the radial distributions of the mid-IR and X-ray surface brightness together with that of the 2MASS J band image smoothed to the mid-IR resolution; as for the X-ray data, we plot a square root of photon counts to show the profile of the plasma density. The directions of the cutting profiles are indicated as A-H in the images of Figure 6; we have selected them in order to avoid apparent point-like sources, which actually deform the shapes of the MIPS 24 $\mu$m contours as seen in the image of NGC~4696 (Fig. 1). The mid-IR profiles show significant excess above the 2MASS (i.e., stellar) distributions at outer radii for NGC~1407 and NGC~4696, while the two look very similar at inner radii for NGC~1395 and NGC~1407. It is clear that the mid-IR emission is stronger in the A, C, F, and H directions (filled circles in Figure 7) than in the B, D, E, and G directions (open circles), which shows a similar trend to the X-ray distribution. The extended mid-IR emission seems to follow the plasma distribution rather than the stellar one, and is therefore likely to be mainly of interstellar dust origin, which contrasts with the central mid-IR emission probably dominated by circumstellar dust. We have confirmed the existence of distributed dust as already predicted by de Jong, et al. (1990) for NGC~4696. 

For the dust to emit observable mid-IR radiation, we might even expect that the dust is heated up to higher temperatures through collisions with plasma electrons in the outer regions of the galaxy where the radiation field heating the dust is dilute. X-ray emission observed with {\it Chandra} is much more extended than the stellar distribution (Fig. 7). In fact, \citet{Gou95} demonstrate that, for NGC~1395, NGC~1407, and NGC~4696, collisional heating by plasma electrons contribute significantly to an increase in the average temperature of diffusely distributed dust at outer radii with their computation. For NGC~4696, \citet{Jon90} find that hot electrons are able to heat the dust to temperatures of about 24 K throughout the whole galaxy, while photons alone heat the dust within $26''$ from the center to an average temperature of about 21 K and within $52''$ to an average temperature of only about 18 K. In addition, as noted by Dwek (1986), the mid-IR to far-IR ratio can be enhanced significantly by stochastic heating of very small grains. Based on the model by Yamada \& Kitayama (2005), we find that the ratio of the 24 um to 160 um brightness ranges from 0.01 to 0.1, depending on the putative size distribution of dust grains (power-law index $p=-2.5\sim-3.5$). Given that the sensitivity in the mid-IR band is about 15 times better than the far-IR band for diffuse sources with the observational parameters as listed in Table 2, this ratio is marginally compatible with the present observations; a typical surface brightness in the outer parts is 0.04 $\sim$ 0.05 MJy/str at 24 $\mu$m and $\lesssim$ 0.8 MJy/str at 160 $\mu$m from Figure 1. Considering the short lifetime against sputtering, such dust must have been supplied quite recently into the ISM of the outer parts of the galaxy, which again favors the merger and evaporation scenario.     

\section{Conclusions}

We have presented mid- to far-IR images and mid-IR spectra of 7 nearby dusty elliptical galaxies observed with the {\it Spitzer} MIPS and IRS in our GO1 program. From past observations, our sample is expected to contain an excessive amount of dust from the viewpoint of its relatively short lifetime against sputtering destruction in hot plasma. As a result, we firmly detect mid-IR emission from warm dust for all the galaxies, far-IR emission from cold dust for 6 galaxies and even spectral emission features from PAHs for 5 out of the 7 galaxies. The central mid-IR component is probably dominated by circumstellar dust emission from evolved stars. We have obtained signatures of interaction between dust and hot plasma; for the emission of the PAHs and the cold dust in the inner parts of the galaxy, both intensities relative to the mid-IR flux decrease with the X-ray luminosity, which can be explained by sputtering destruction. The unusually high ratio of the 11.3 $\mu$m to the PAH emission features shortward of 10 $\mu$m observed for our sample might reflect that the PAHs are truly under sputtering destruction. The far-IR emission becomes colder as the far-IR flux is lower, which can be explained by the shorter lifetime of smaller grains under sputtering destruction.
We have found a correlation between the two components, which indicates that the origins of the excess cold dust and the PAHs are the same. Since the presence of the PAHs is thought to be a sign of a recent merger, the excess dust is also likely to be brought in by the same merger event. For the dust emission in the outer parts of the galaxy, the extended mid-IR emission appear to follow the plasma distribution rather than the stellar distribution for the three X-ray-brightest galaxies in our sample. This spatial correspondence between mid-IR and X-ray not only indicate the existence of diffusely-distributed interstellar dust, but may also suggest the dominance of hot electron collision over stellar radiation for stochastic heating of such distributed dust. Our observational results consistently favor the merger and evaporation scenario rather than the cooling flow scenario for the origin of the excess dust.
 
With the advent of {\it Spitzer}, we can now perform detailed image and spectroscopy studies of dust in hot plasma of elliptical galaxies. Following {\it Spitzer}, {\it AKARI} (ASTRO-F; Murakami 2006), the Japanese infrared astronomical satellite, has been launched successfully in February, 2006, with which we plan to observe these elliptical galaxies. {\it AKARI} has unique capabilities such as 4-far-IR-band photometry and near-IR spectroscopy that enable us to unambiguously determine the temperature and mass of the cold dust and to diagnose the ionization state of the PAHs, respectively.

\bigskip

This work is based on observations made with the {\it Spitzer} Space Telescope, which is operated by the Jet Propulsion Laboratory, California Institute of Technology under NASA contract 1407. 
We would express our gratitude to the IRS/MIPS team and the SSC for their dedicated work in generating the BCD and developing data analysis tools. This work is financially supported by a Grant-in-Aid from the Ministry of Education, Culture, Sports, Science and Technology in Japan (No. 17740123).

\clearpage

\begin{figure}
\begin{center}
\FigureFile(43mm,43mm){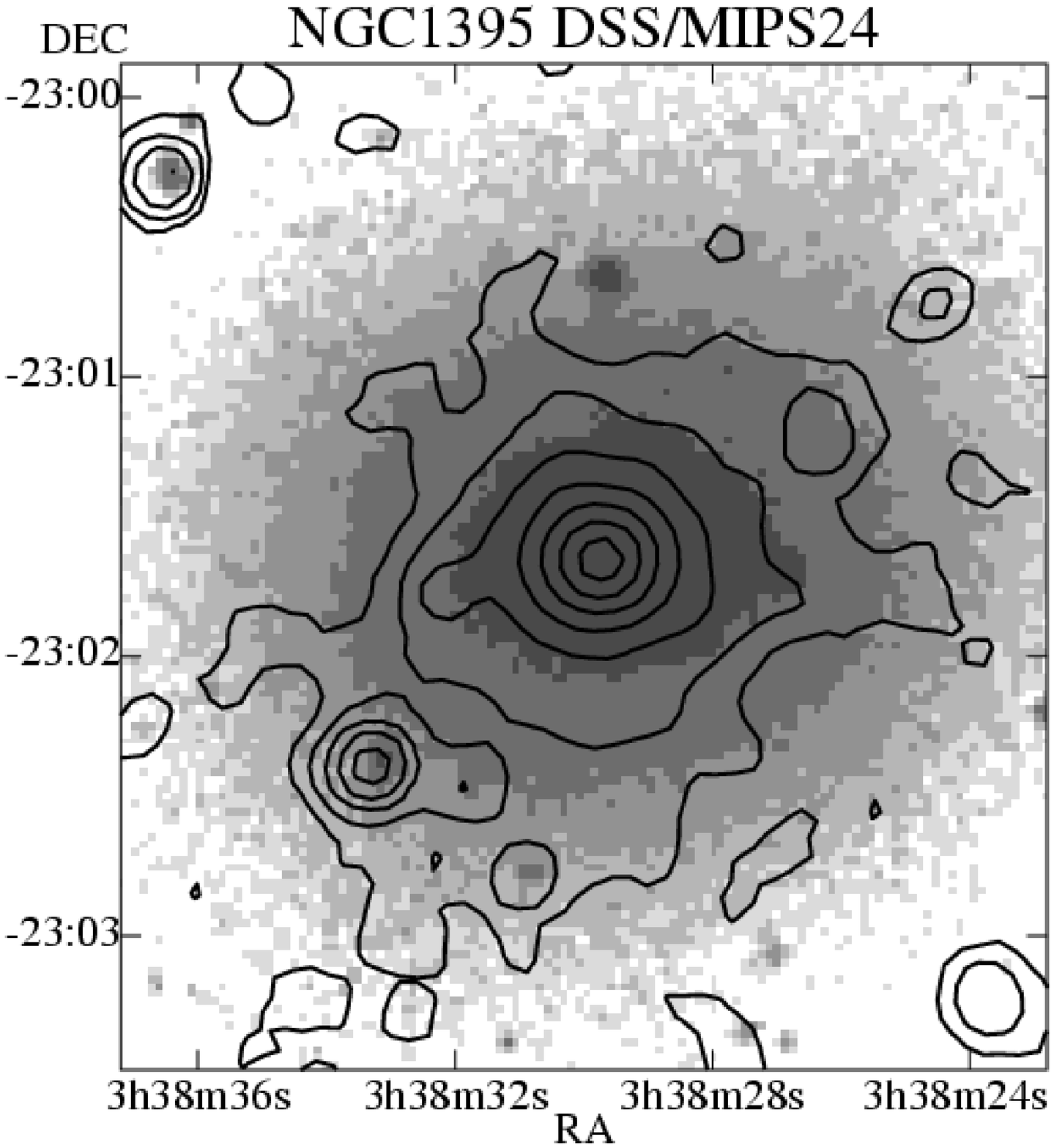}\FigureFile(43mm,43mm){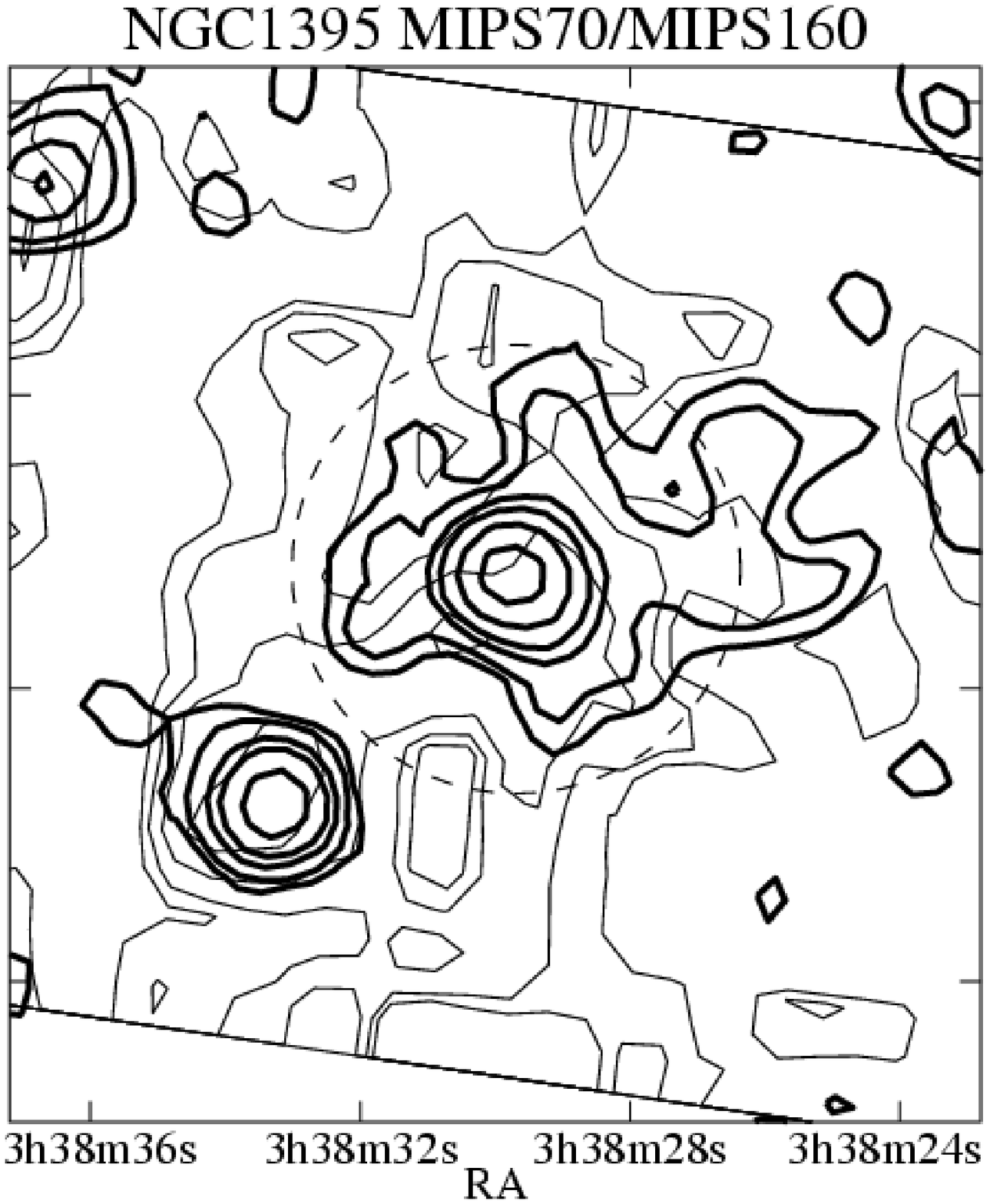}\FigureFile(43mm,43mm){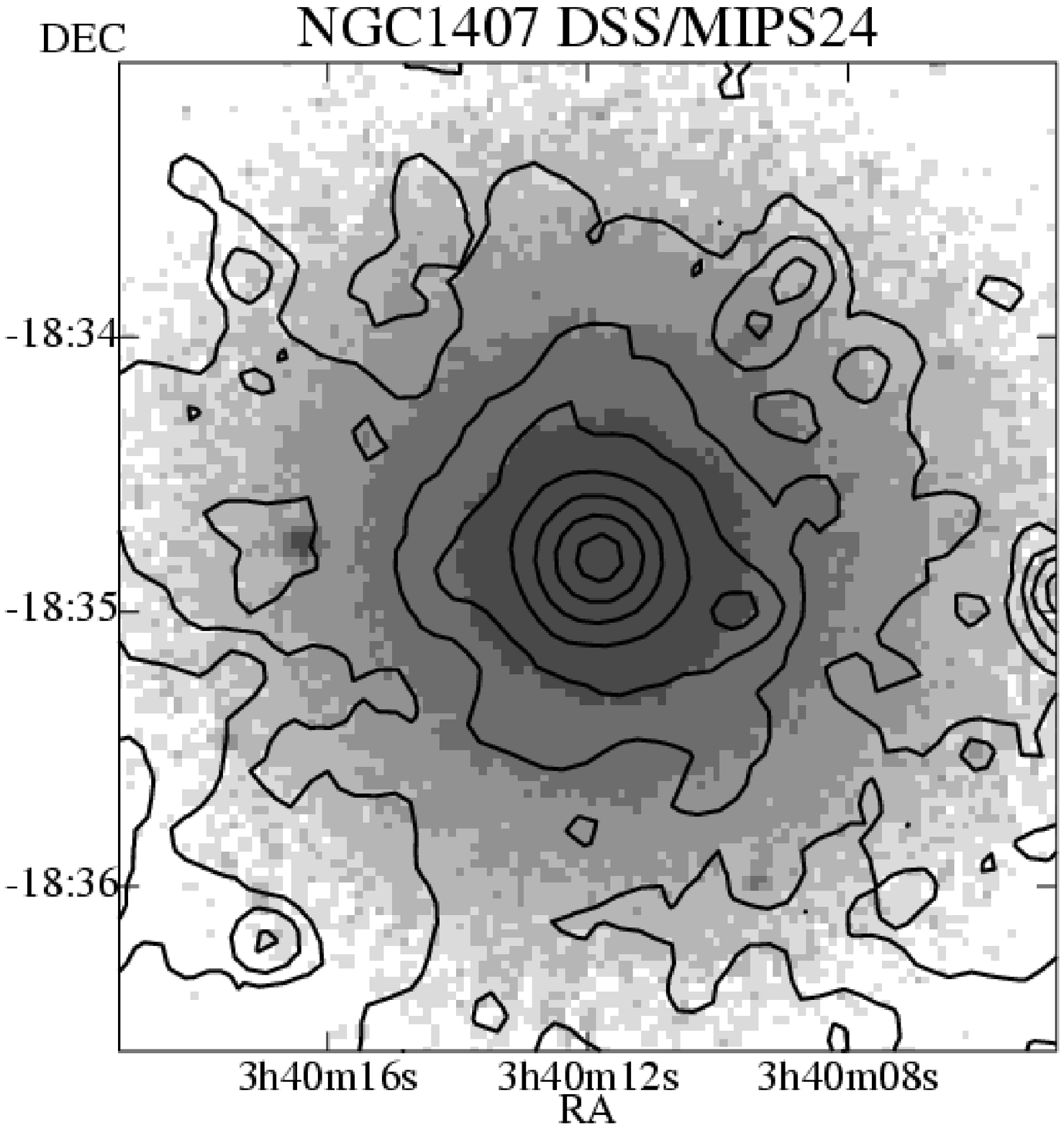}\FigureFile(43mm,43mm){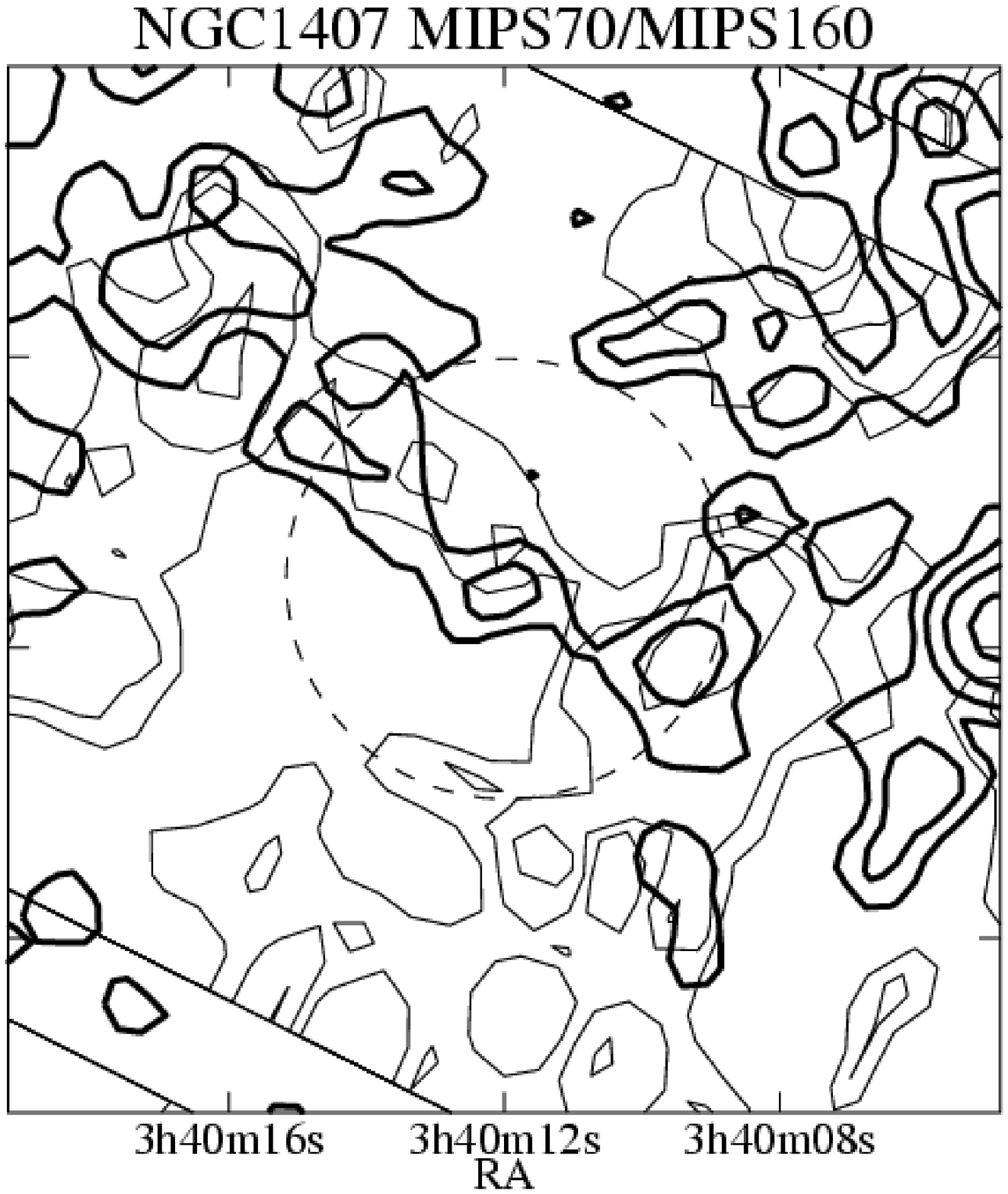}\\
\FigureFile(43mm,43mm){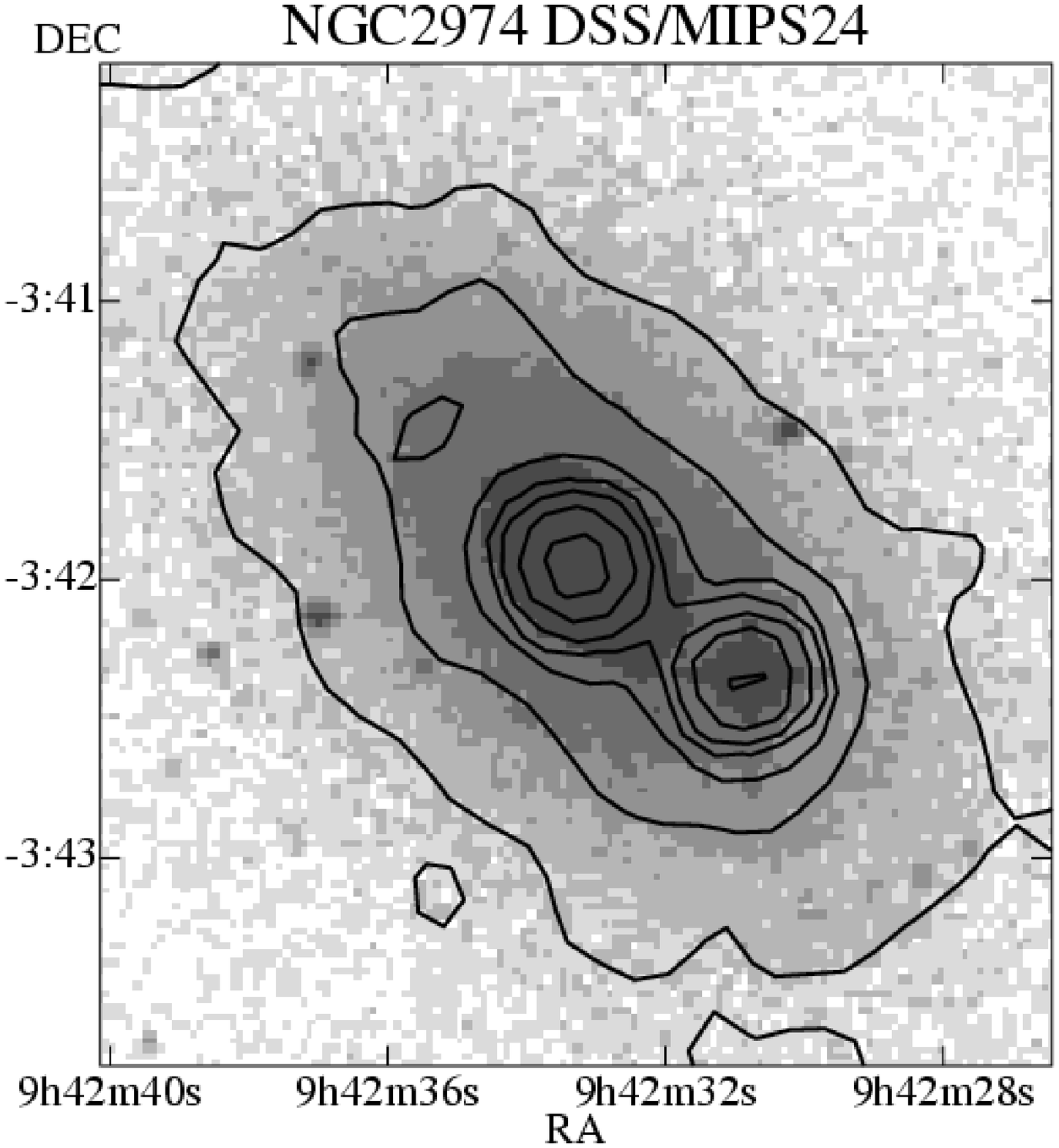}\FigureFile(43mm,43mm){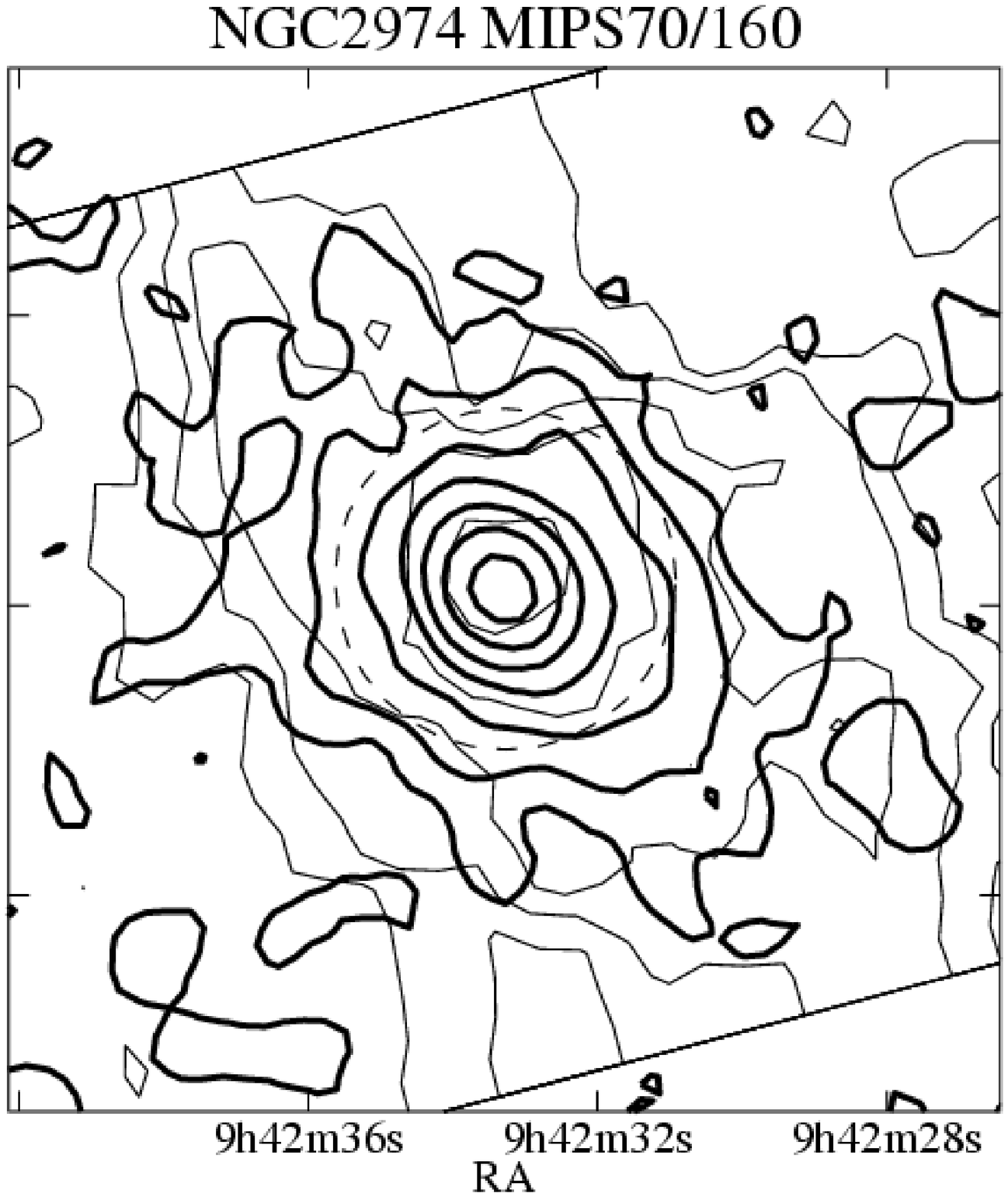}\FigureFile(43mm,43mm){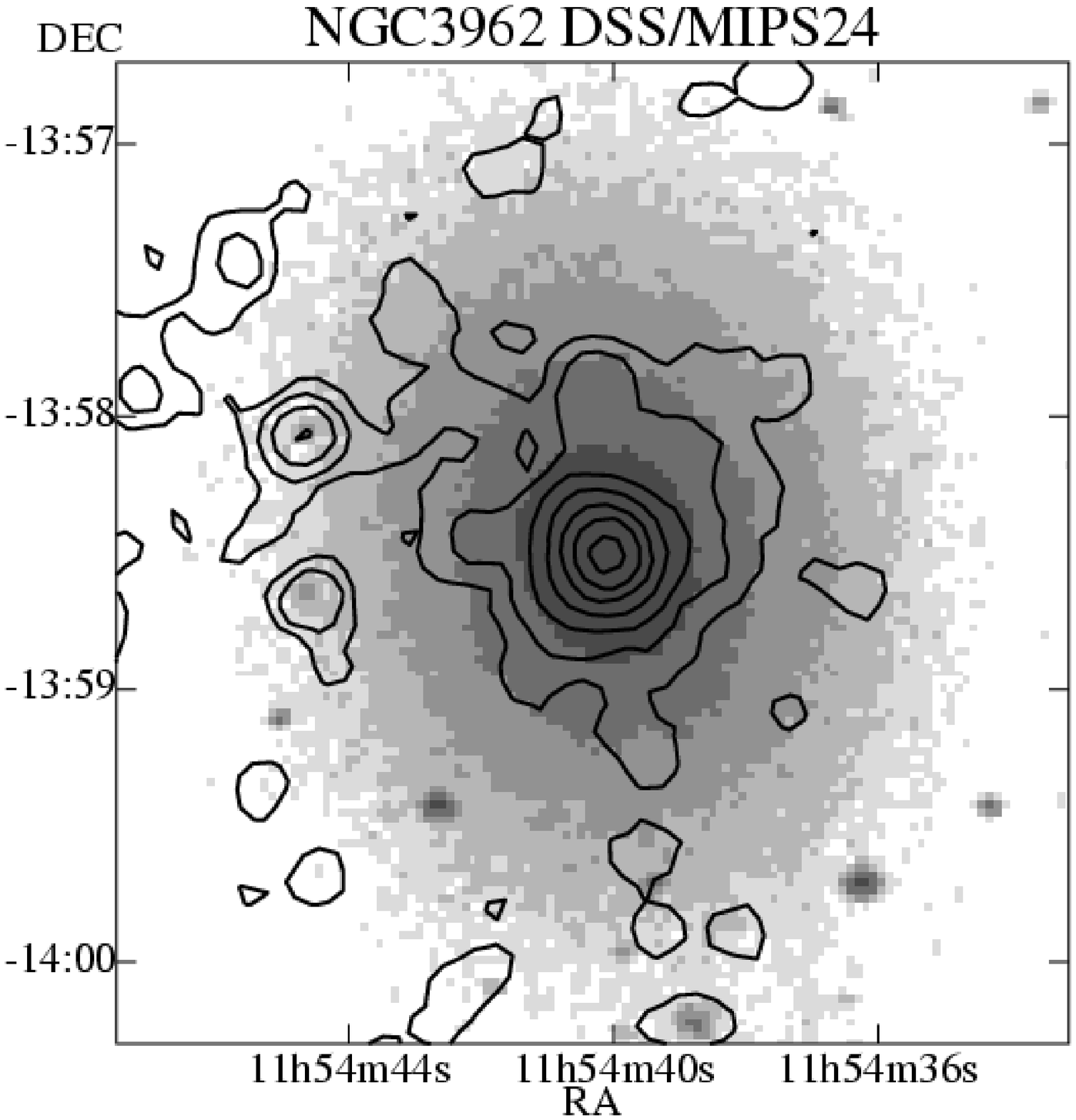}\FigureFile(43mm,43mm){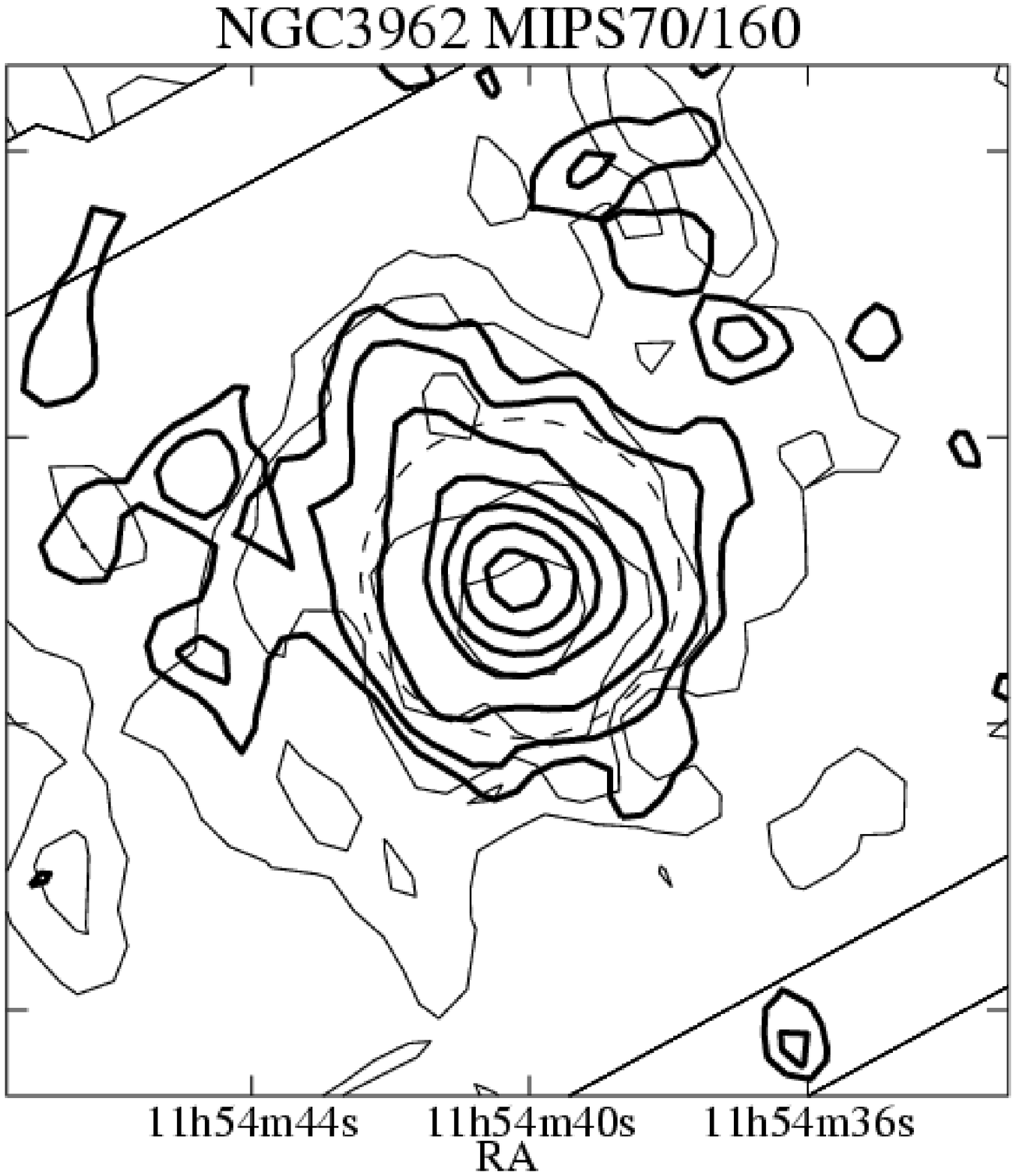}\\
\FigureFile(43mm,43mm){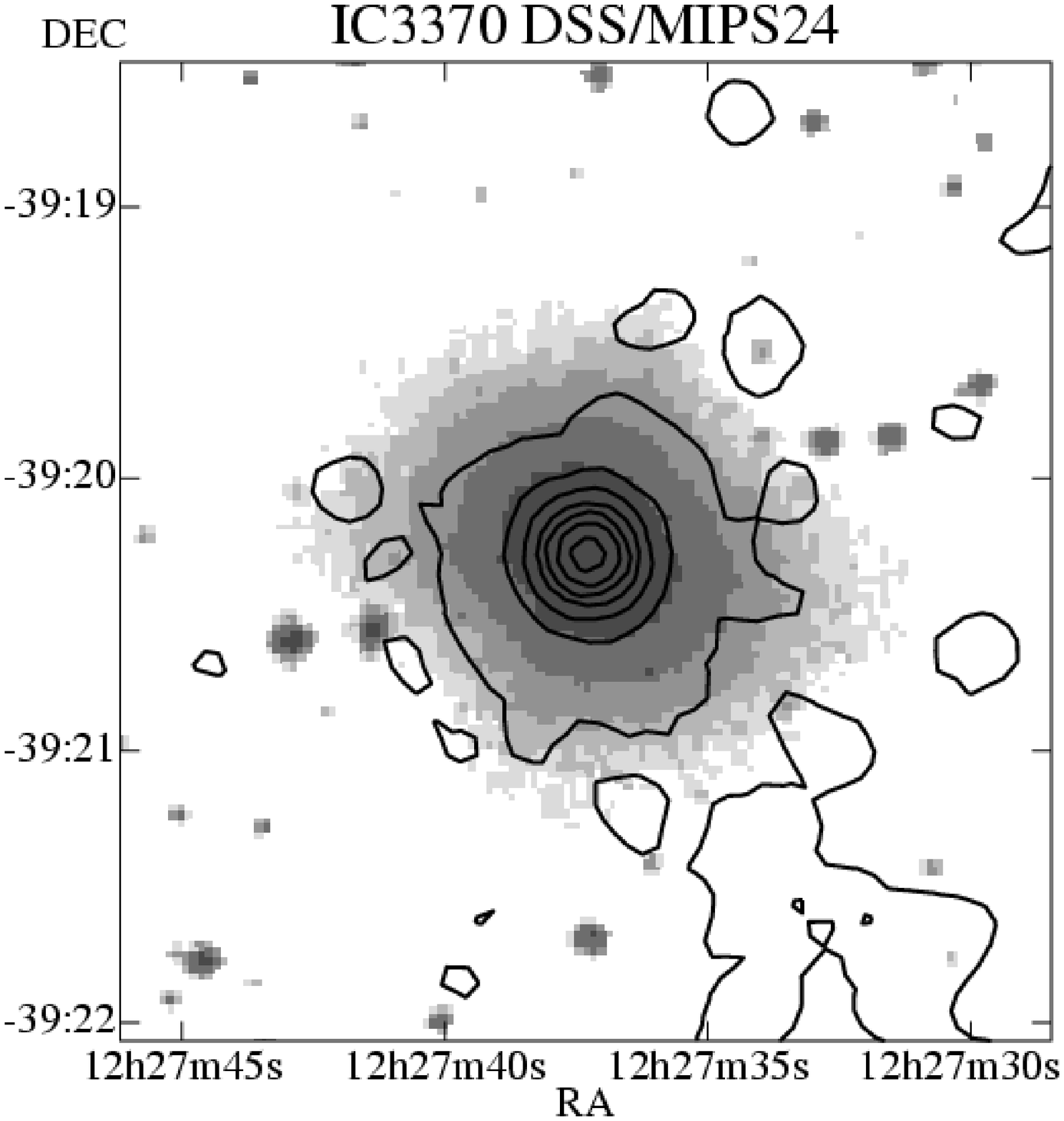}\FigureFile(43mm,43mm){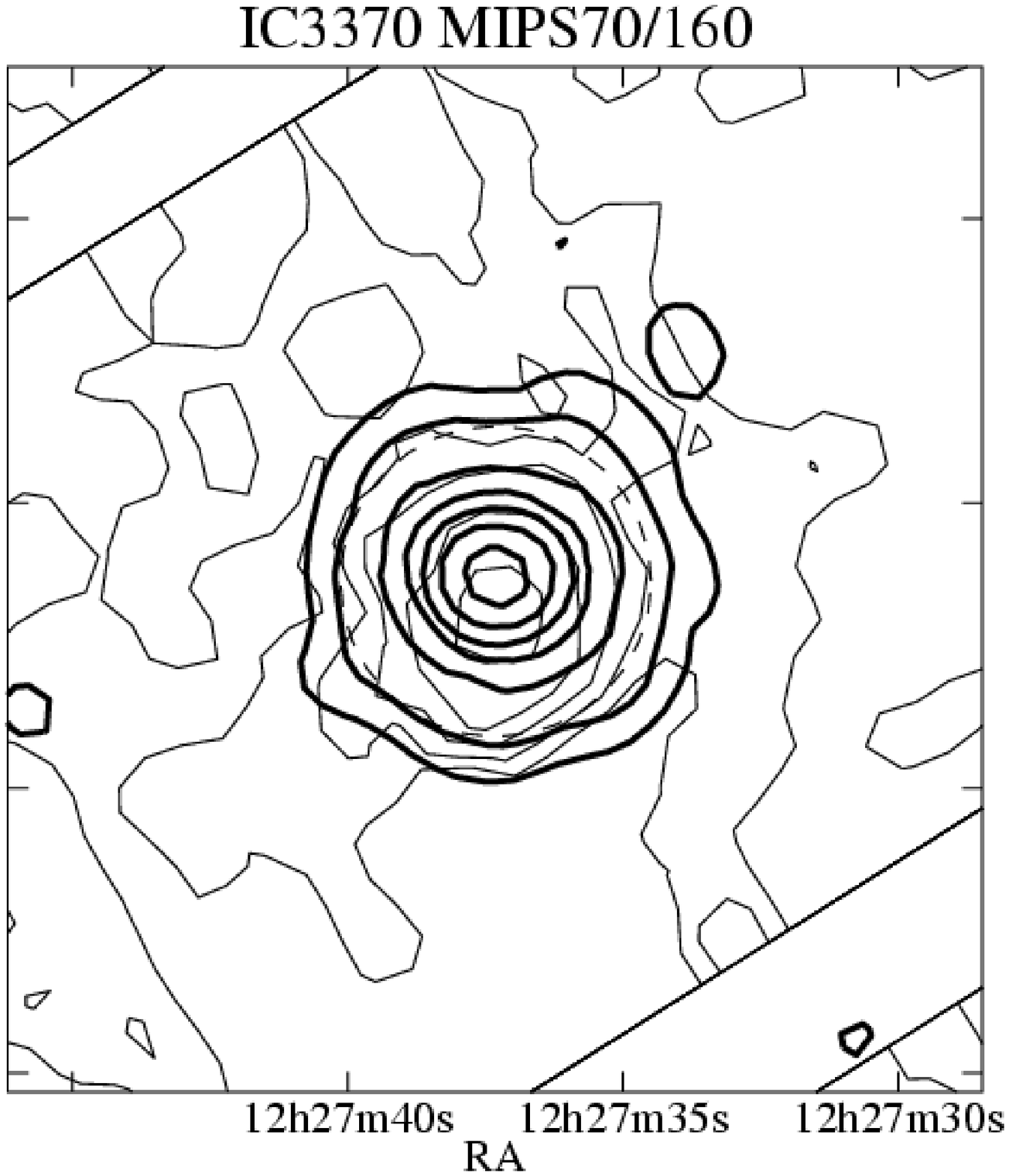}\FigureFile(43mm,43mm){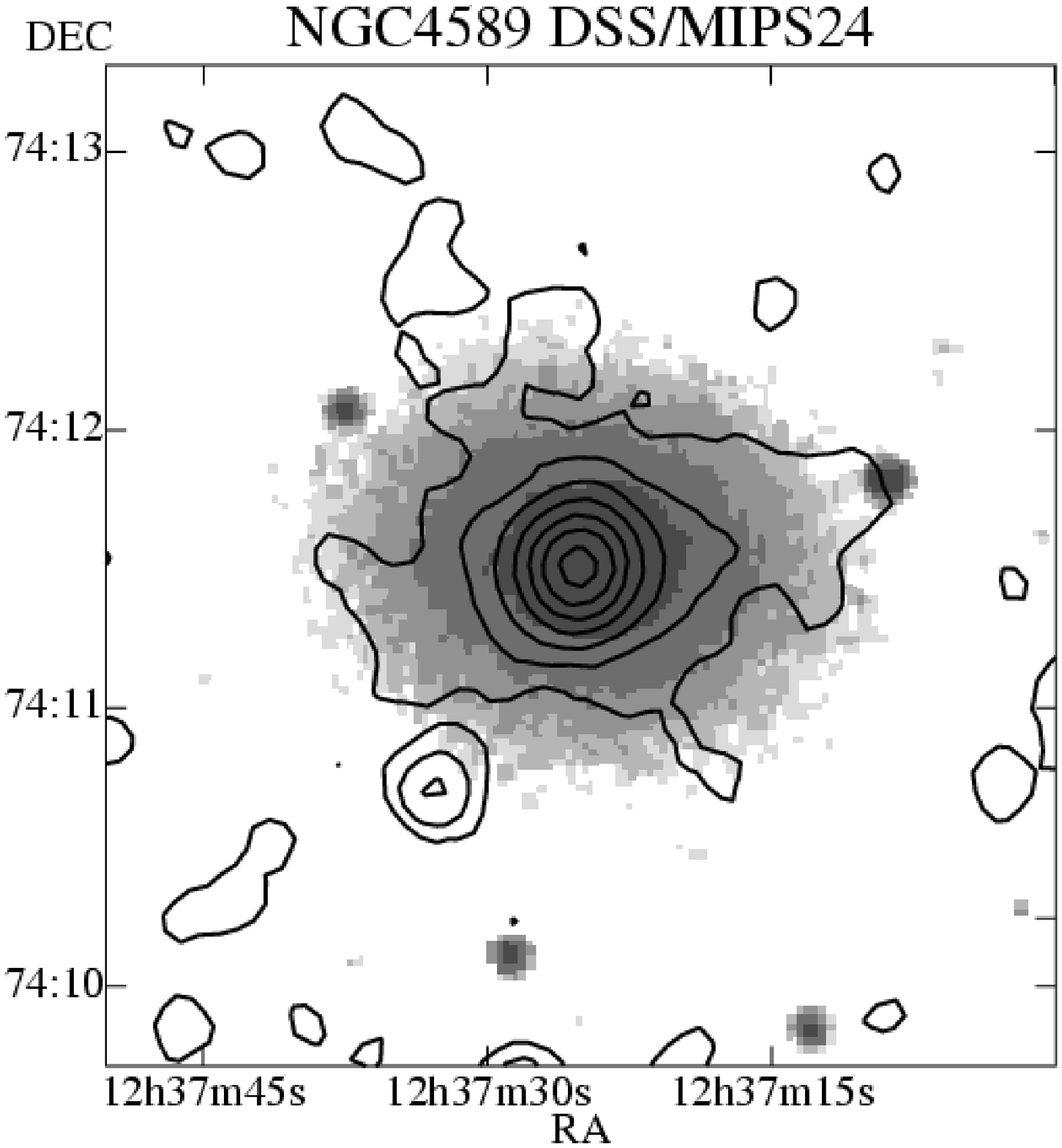}\FigureFile(43mm,43mm){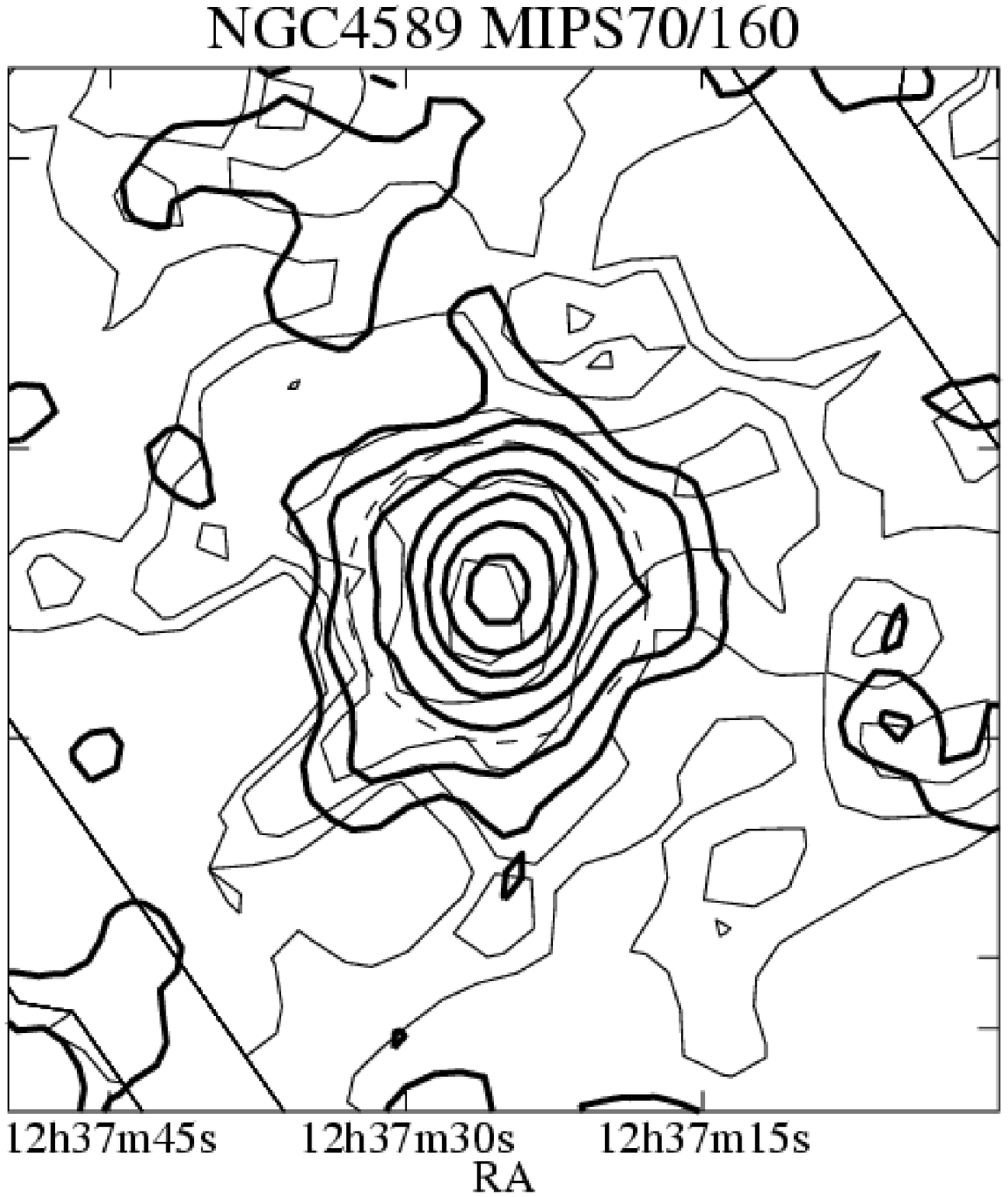}\\
\FigureFile(43mm,43mm){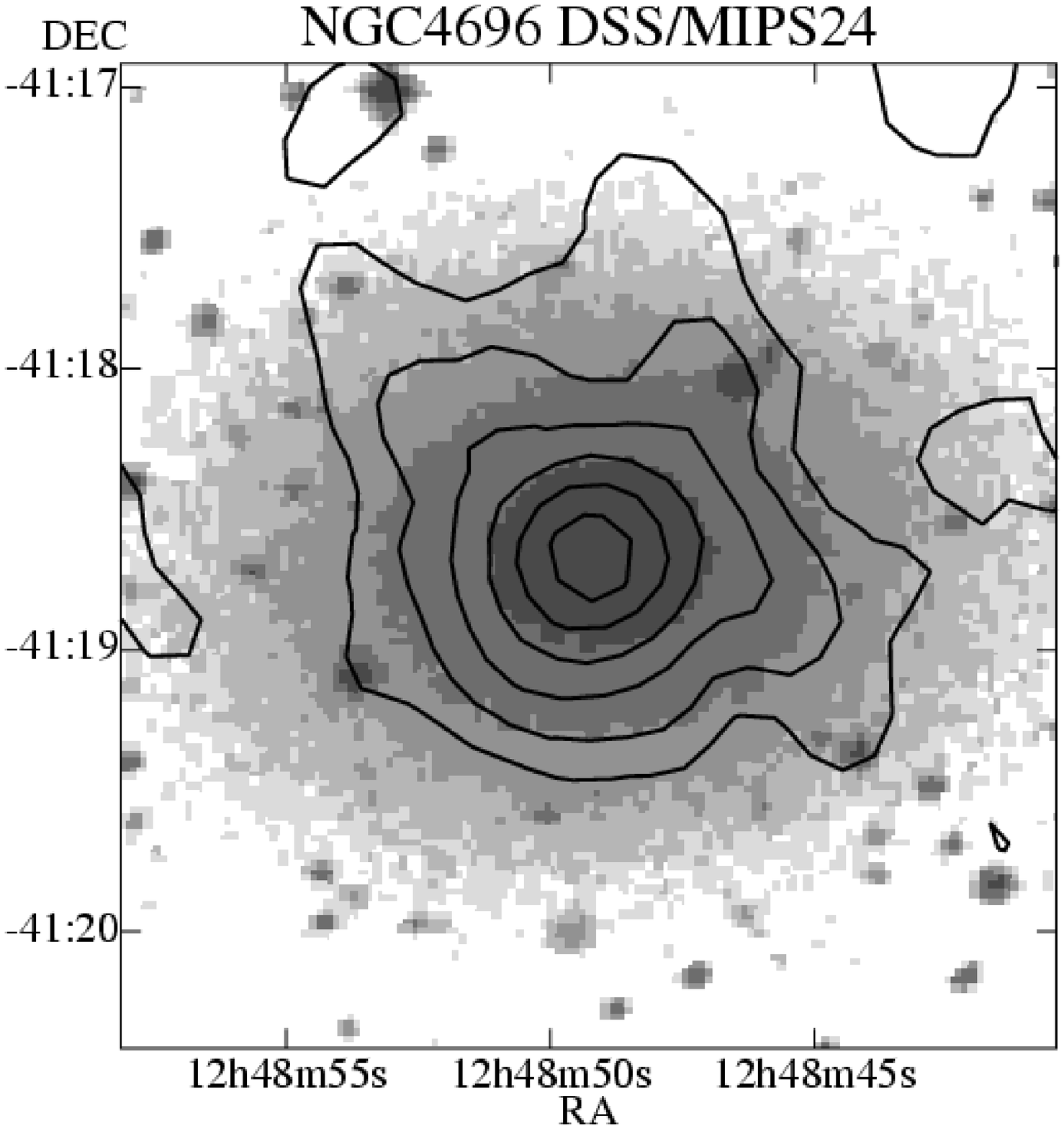}\FigureFile(43mm,43mm){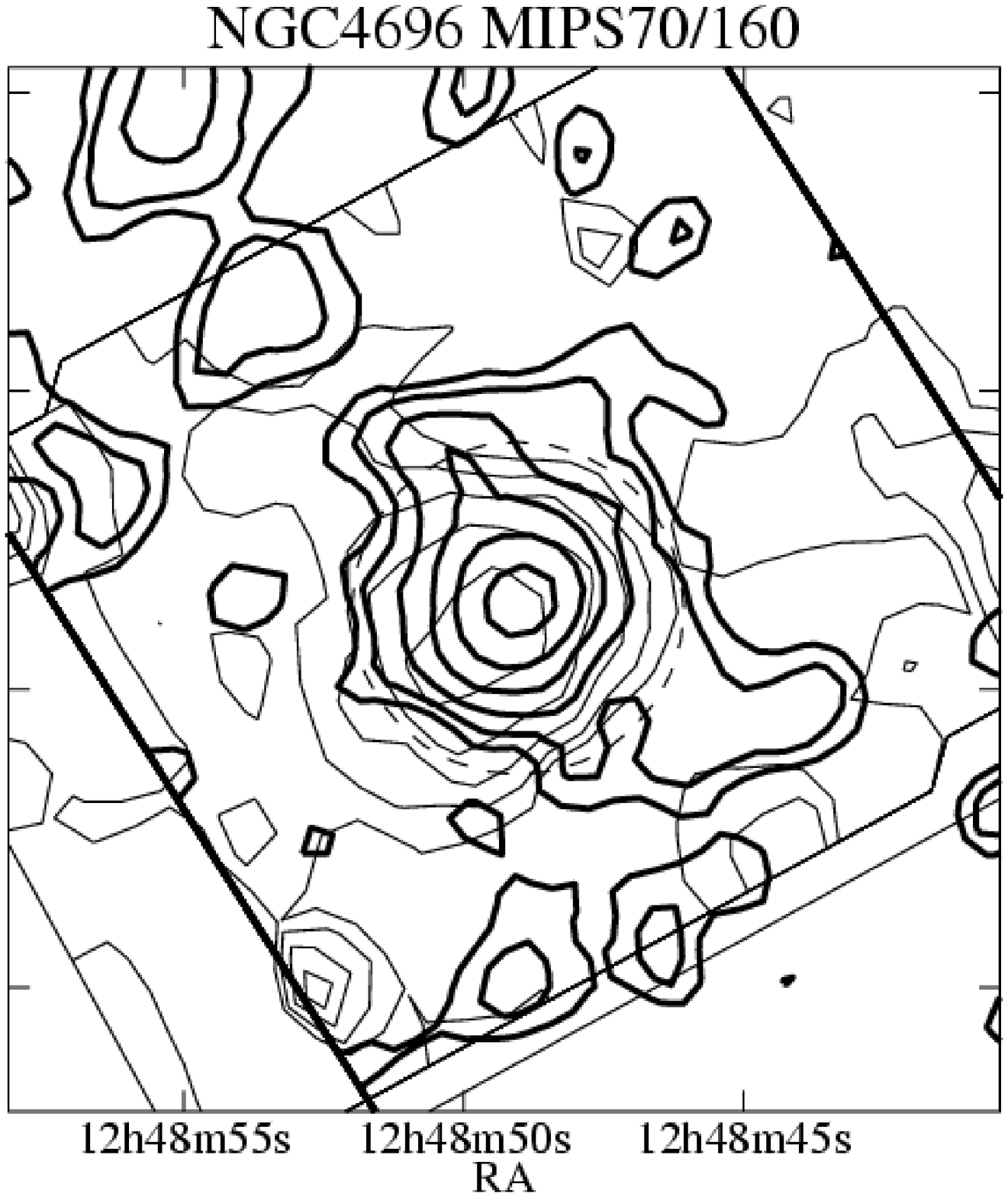}
\end{center}
\end{figure}
\clearpage

\begin{figure}
\caption{MIPS mosaic images. For each galaxy, the MIPS 24 $\mu$m contours are overlaid on the DSS optical image in the left panel, while the MIPS 70 $\mu$m contours (thick lines) are overlaid on the MIPS 160 $\mu$m contours (thin lines) in the right panel. For MIPS 24 $\mu$m, contours are spaced at 80, 50, 30, 18, 10, 5, and 2 \% levels of the background-subtracted peak surface brightness of 2.2 (NGC~1395), 1.8 (NGC~1407), 1.8 (NGC~3962), 2.7 (NGC~2974), 4.7 (IC~3370), and 1.7 MJy/str (NGC~4589), and 80, 50, 30, 18, 10, and 5 \% levels of 0.6 MJy/str (NGC~4696). For MIPS 70 $\mu$m, contour levels are 80, 50, 30, 18, 10, and 5 \% of 3.3 MJy/str (NGC~1395), 80, 50, 30, and 18 \% of 1.5 MJy/str (NGC~1407), and 80, 50, 30, 18, 10, 5, and 2 \% of 22.9 (NGC~2974), 12.9 (NGC~3962), 31.0 (IC~3370), 9.7 (NGC~4589), and 5.5 MJy/str (NGC~4696). For MIPS 160 $\mu$m, contour levels are 80, 50, 30, and 18 \% of 3.2 (NGC~1395), 4.5 (NGC~1407), and 6.3 MJy/str (NGC~3962), and 80, 50, 30, 18, and 10 \% of 15.2 (NGC~2974), 14.9 (IC~3370), 6.1 (NGC~4589) and 4.6 MJy/str (NGC~4696). Typical 1-sigma noise levels are 0.01 MJy/str, 0.05 MJy/str, and 0.1 MJy/str for the MIPS 24 $\mu$m, 70 $\mu$m, and 160 $\mu$m band images, respectively. The dashed circles correspond to photometry apertures, for which the flux densities in Table 3 are derived.}
\end{figure}    

\clearpage

\begin{figure}
\begin{center}
\FigureFile(120mm,120mm){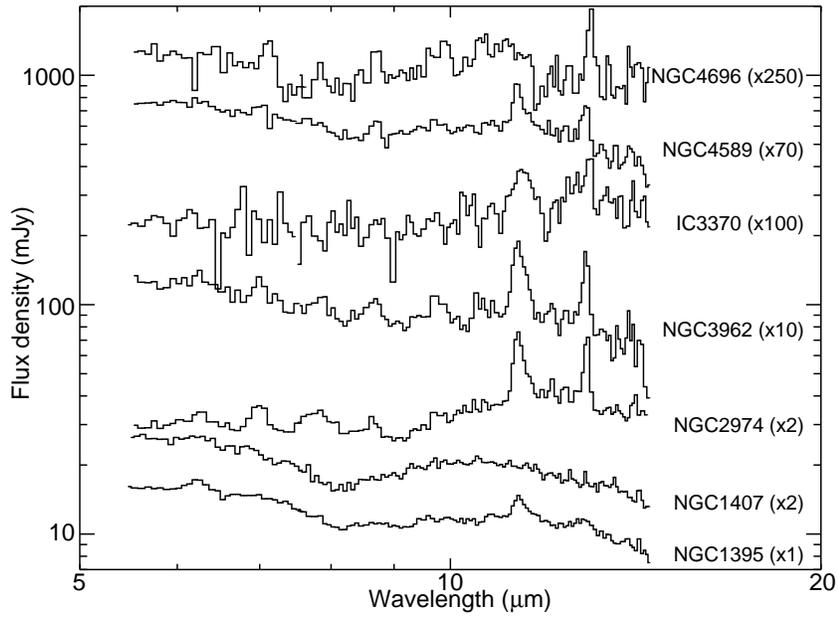}
\end{center}
\caption{IRS/SL spectra. Each spectrum is labeled with the object name and the gain factor applied for clarity of presentation. The 12.7 $\mu$m emission feature may be contaminated by the [NeII] 12.81 $\mu$m fine-structure line. }
\end{figure}

\clearpage

\begin{figure}
\begin{center}
\FigureFile(120mm,120mm){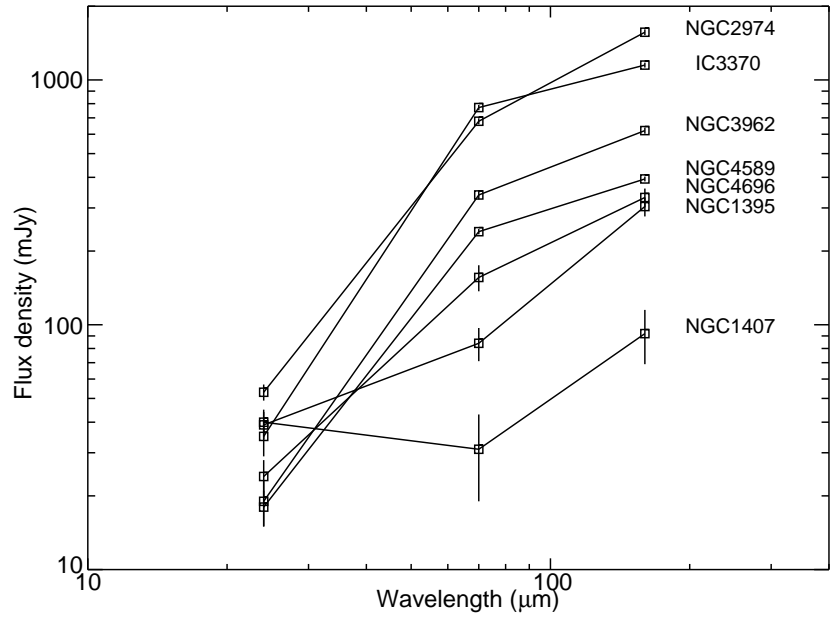}
\end{center}
\caption{Spectral energy distributions obtained with the MIPS 3 bands. A blackbody spectrum (emissivity law $\propto$ $\lambda^{-1}$) with a temperature of 30 K is plotted together.}
\end{figure}

\clearpage

\begin{figure}
\begin{center}
\FigureFile(80mm,80mm){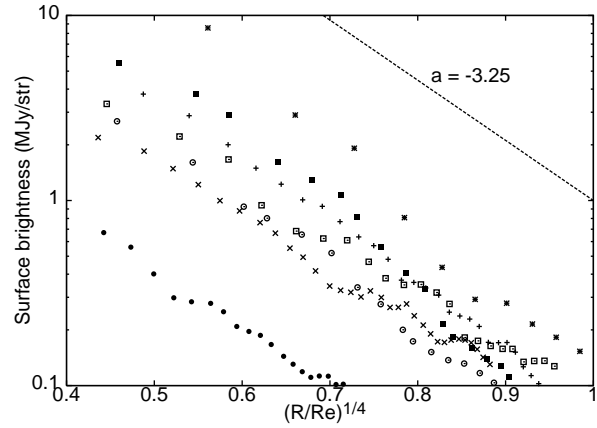}
\end{center}
\caption{MIPS 24 $\mu$m surface brightness profiles of our sample galaxies as a function of the semimajor axis normalized by $R_e$ to the 1/4 power, where $R_e$ is the effective radius. Different symbols indicate the profiles of different galaxies. A de Vaucouleurs profile should be a straight line with a slope of $-3.25$ in this plot.}
\end{figure}

\clearpage
\begin{figure}
\begin{center}
\FigureFile(70mm,70mm){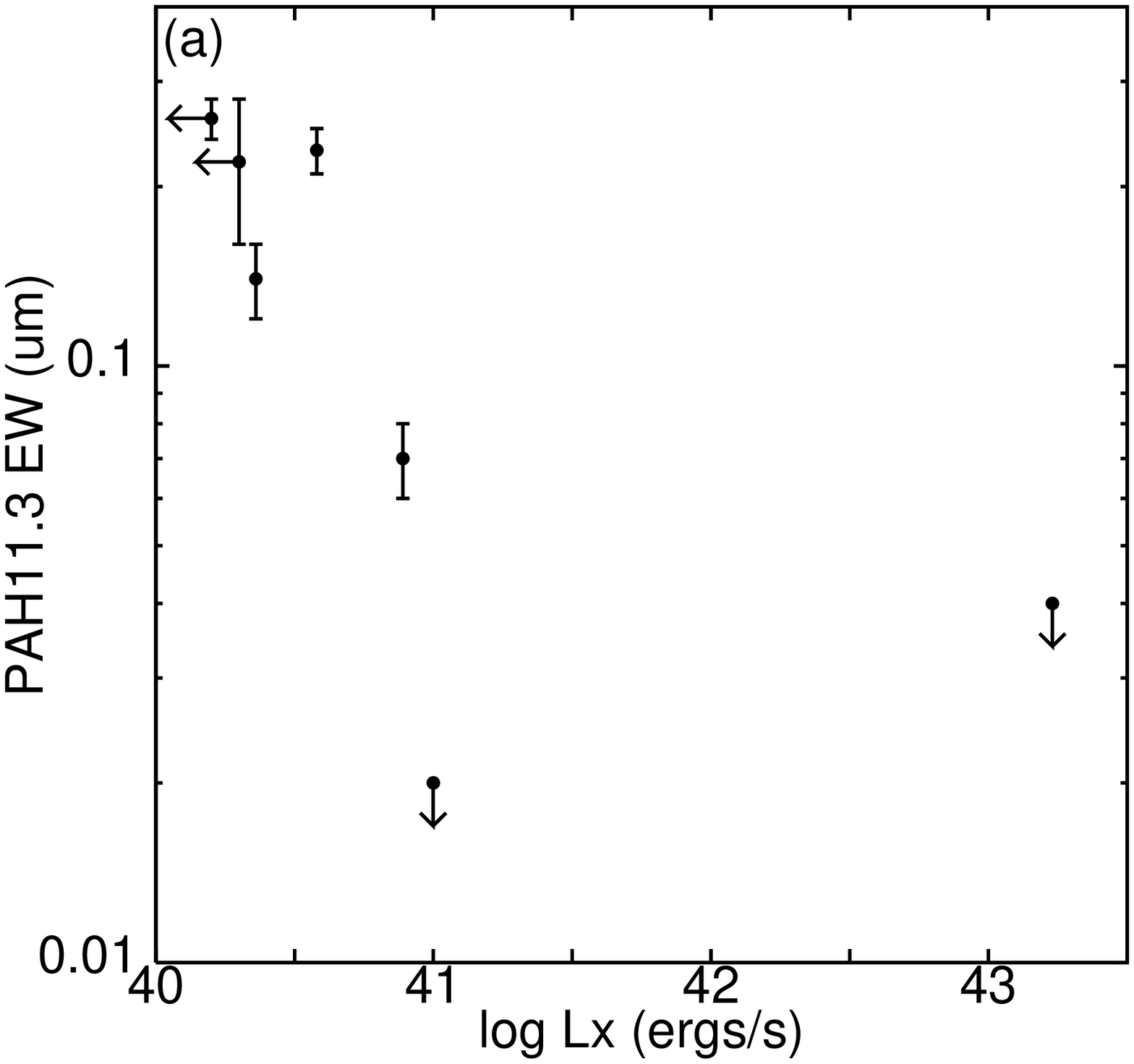}\FigureFile(70mm,70mm){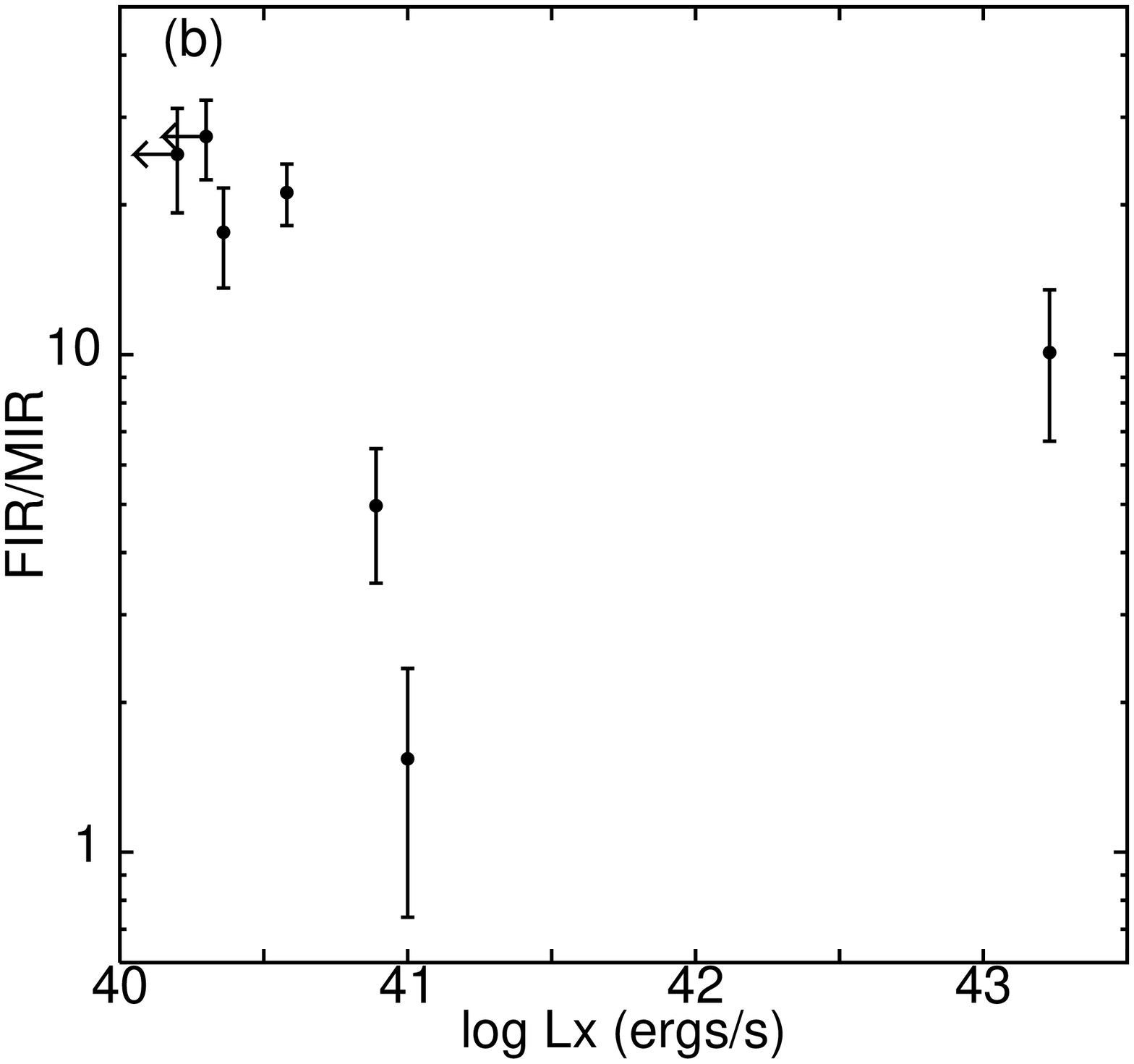}\\
\FigureFile(70mm,70mm){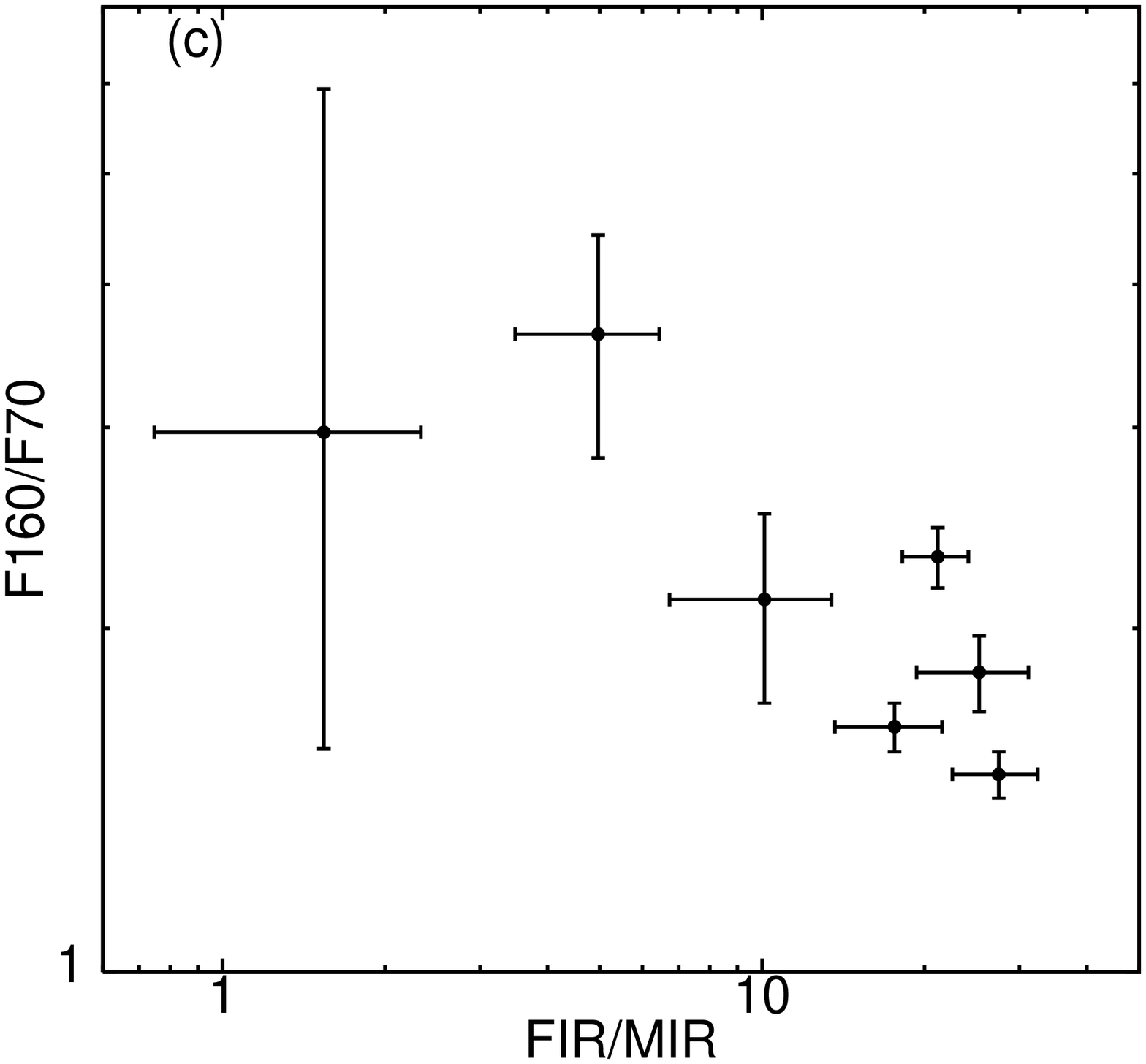}\FigureFile(70mm,70mm){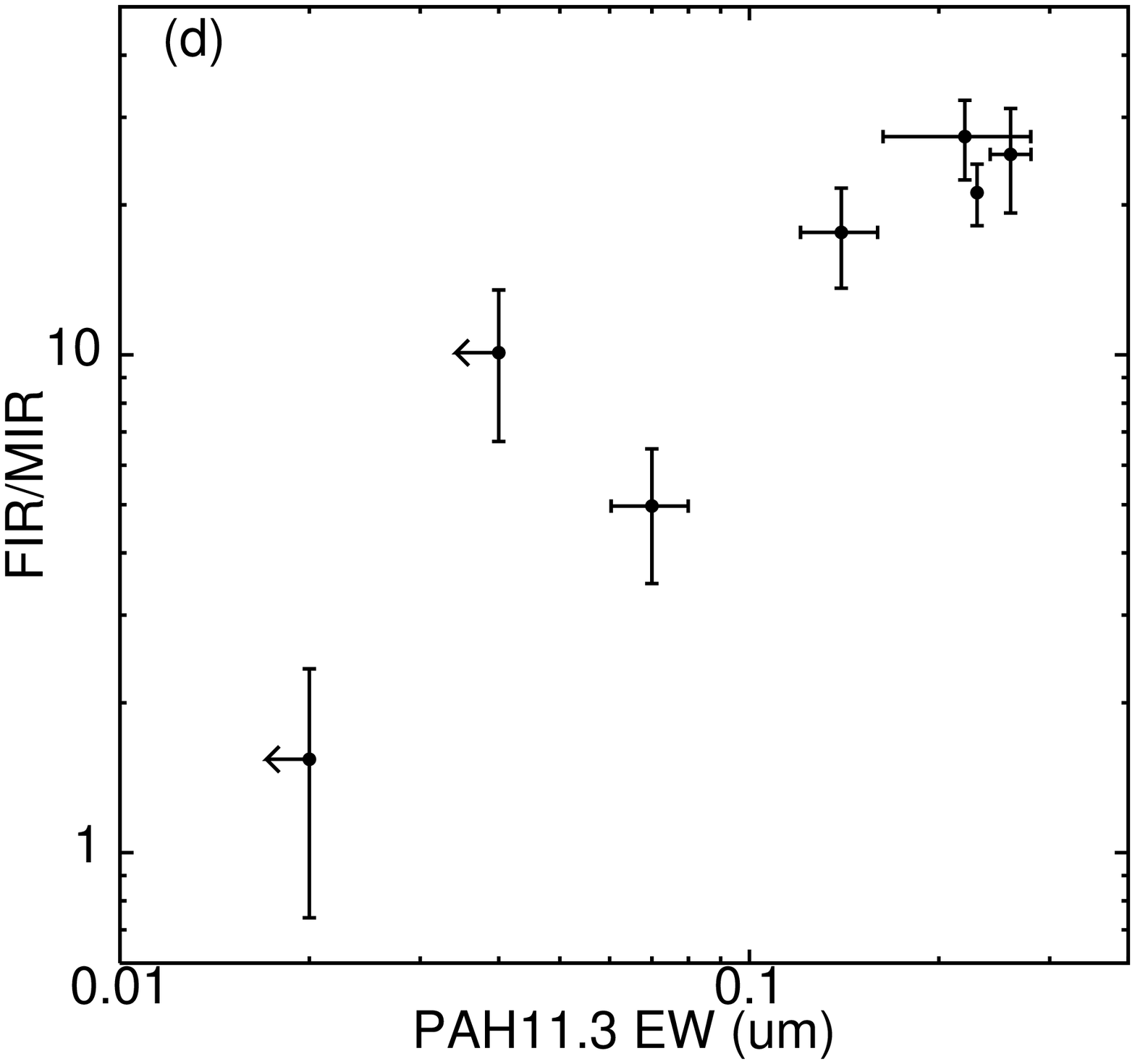}
\end{center}
\caption{Scatter plots between the X-ray luminosity, the equivalent width of the 11.3 $\mu$m PAH emission feature, the ratio of the average of the MIPS 70 $\mu$m and 160 $\mu$m fluxes (FIR) to the MIPS 24 $\mu$m flux (MIR), and the ratio of the MIPS 70 $\mu$m (F70) to the 160 $\mu$m flux (F160).}
\end{figure}

\clearpage

\begin{figure}
\begin{center}
\FigureFile(60mm,60mm){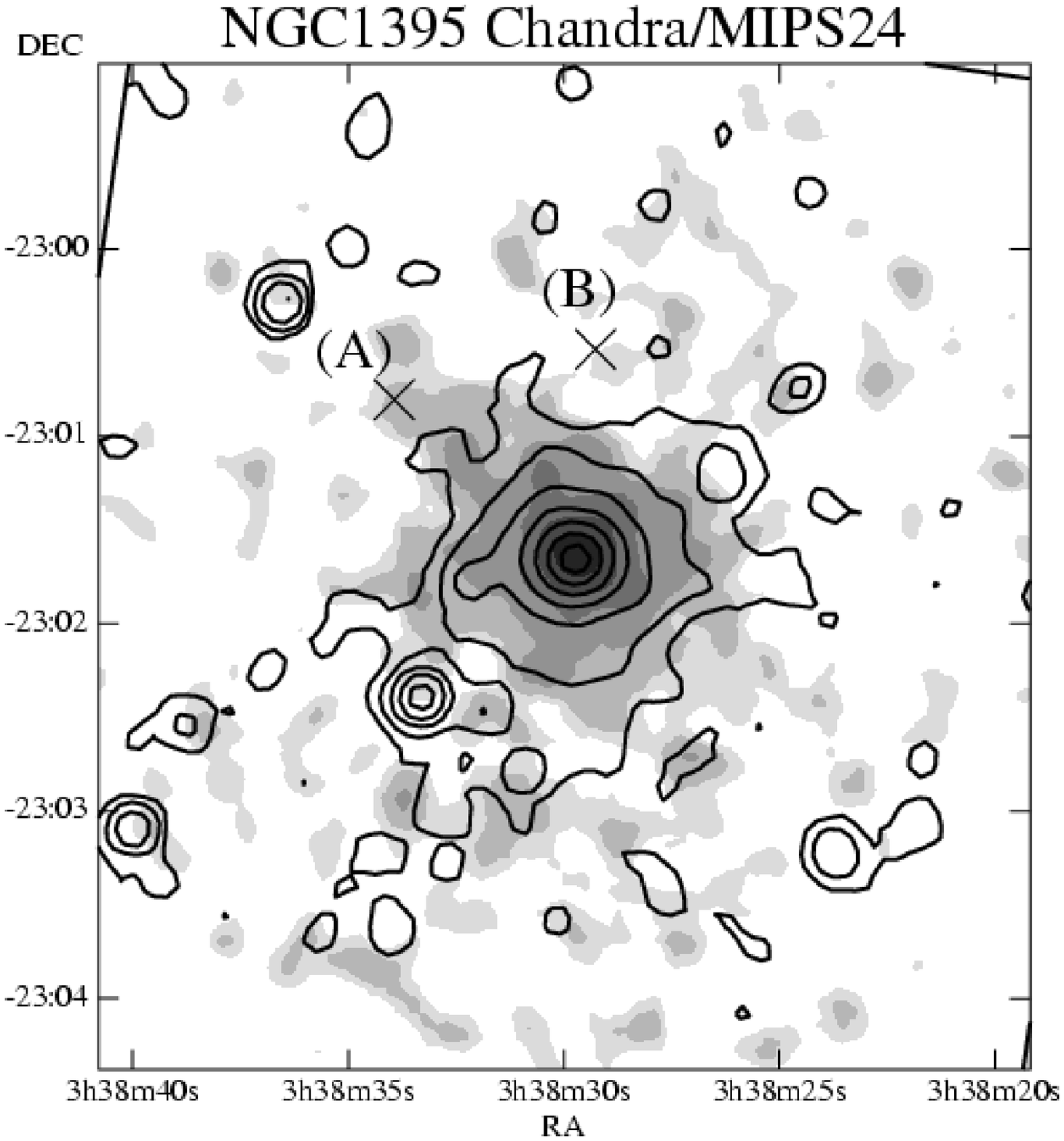}\FigureFile(60mm,60mm){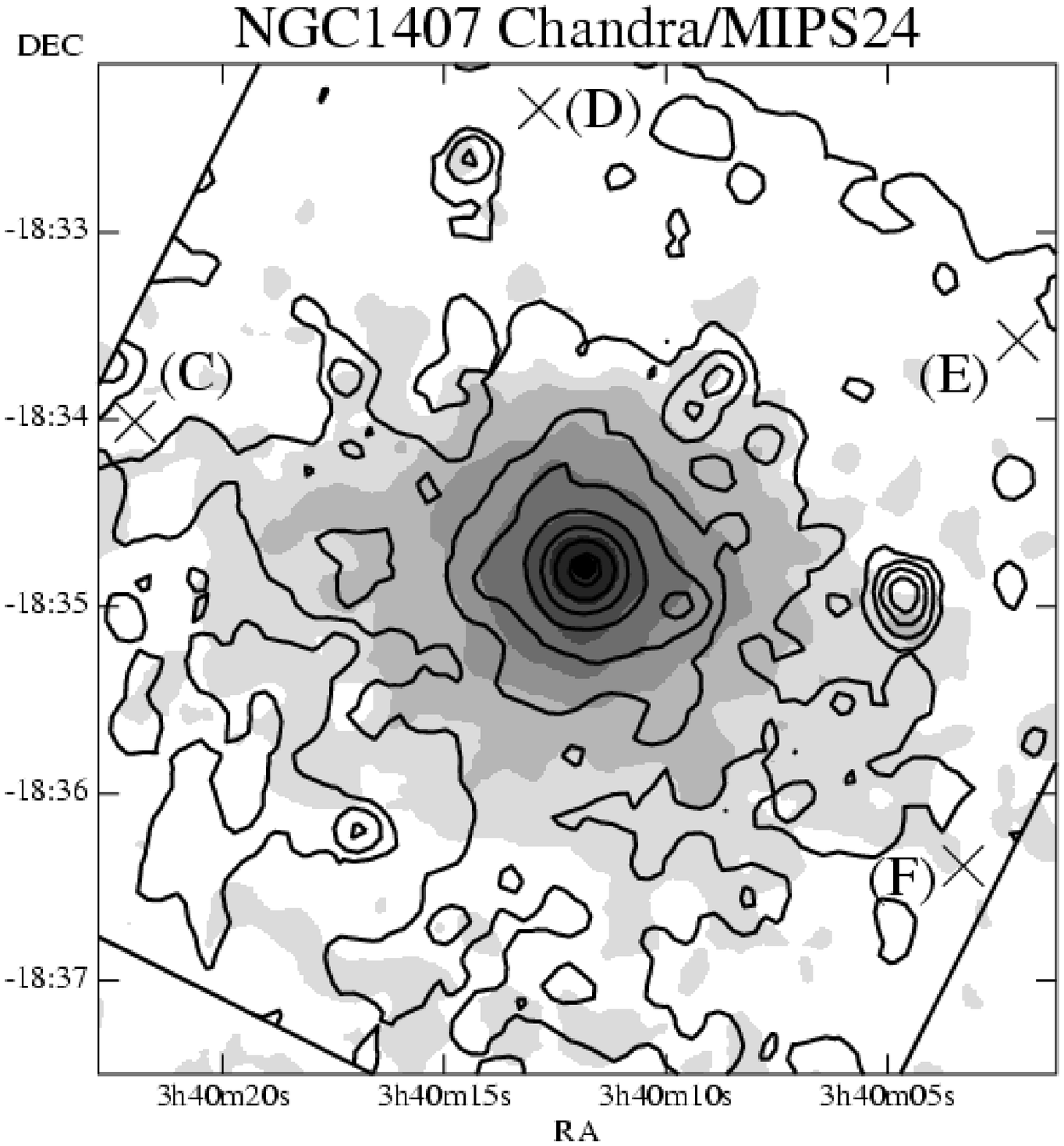}\FigureFile(38.3mm,38.3mm){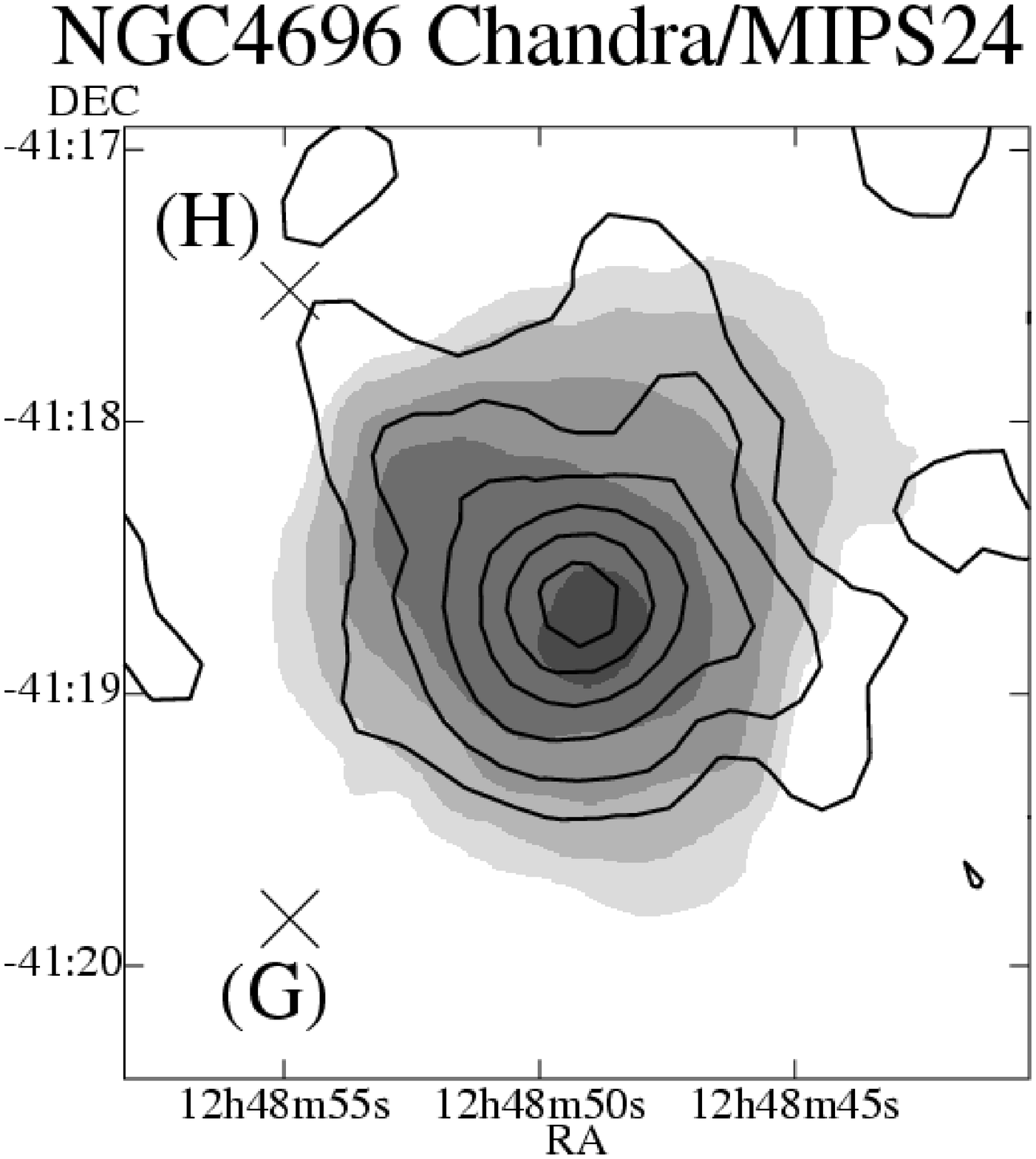}
\end{center}
\caption{MIPS 24 $\mu$m images are overlaid on {\it Chandra} X-ray smoothed images for NGC~1395, NGC~1407, and NGC~4696. The X-ray images are plotted in logarithmic scale, while the MIPS 24 $\mu$m contours are spaced at the same levels as in Figure 1. Point sources are removed from the X-ray images. The positions (A)-(H) in the images are used to create the radial plots in Figure 7.}
\end{figure}

\clearpage

\begin{figure}
\begin{center}
\FigureFile(160mm,160mm){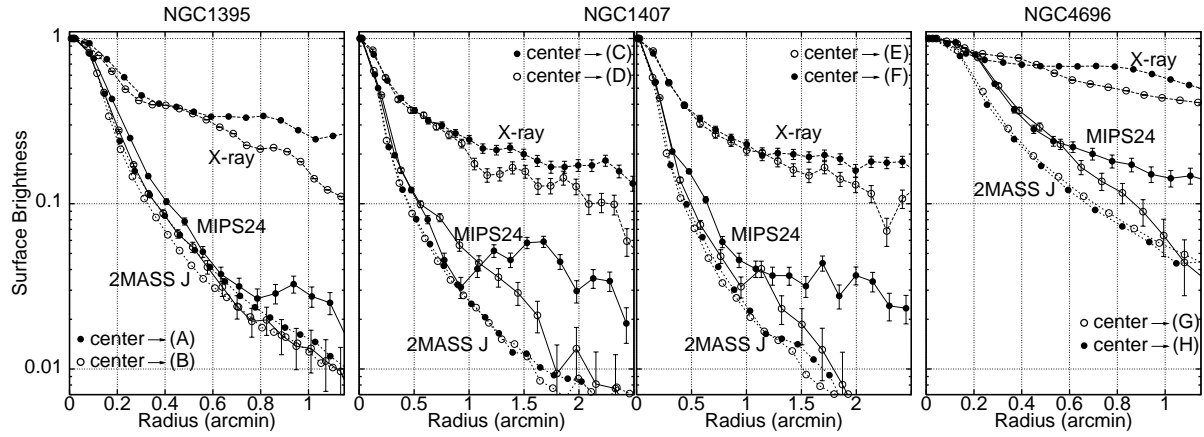}
\end{center}
\caption{Radial distributions of the surface brightness in the X-ray ({\it Chandra} ACIS 0.3-10 keV), mid-IR (MIPS 24 $\mu$m), and near-IR (2MASS J) bands along the directions from the center of the galaxy to the positions (A)-(H) in Figure 6, with each normalized to the peak. Note that, as for the X-ray data, a square root of photon counts per bin is plotted.}
\end{figure}

\clearpage

\begin{table*}
\caption{Properties of the Observed Galaxies}
\begin{center}
\begin{tabular}{lrrrrrr}
\hline\hline
Name & Type & D & $D_{n}$\footnotemark[$*$]&Log$L_{\rm B}$\footnotemark[$\dagger$] & Log$L_{\rm IR}$\footnotemark[$\ddagger$] & Log$L_{\rm X}$\footnotemark[$\S$]\\
 & & (Mpc) & (arcsec) & (L$_{\odot}$) & (L$_{\odot}$) & (ergs s$^{-1}$) \\ 
\hline
NGC~1395 & E2  & 20.5 & 69 & 10.36 & 8.45 & 40.89 \\
NGC~1407 & E0  & 20.6 & 69 & 10.58 & 8.64 & 41.00 \\
NGC~2974 & E4  & 28.3 & 53 & 10.29 & 9.27 & 40.58 \\
NGC~3962 & E1  & 21.7 & 50 & 10.13 & 8.54 & $<$40.22 \\
IC~3370  & E2  & 39.1 & 49 & 10.69 & 9.85 &$\dots$ \\
NGC~4589 & E2  & 24.6 & 46 & 10.33 & 8.96 & 40.36 \\
NGC~4696 & E2  & 37.0 & 47 & 10.83 & 9.36 & 43.23 \\
\hline
\\
\multicolumn{7}{@{}l@{}}{\hbox to 0pt{\parbox{160mm}{\footnotesize

\par\noindent
\footnotemark[$*$] Optical angular diameter given in Table 1 of Faber et al. (1989).  
\par\noindent
\footnotemark[$\dagger$] Total blue luminosity of the galaxies, derived using a solar absolute magnitude M$_{\rm bol,\odot} = +4.75$. 
\par\noindent
\footnotemark[$\ddagger$] The IR (1$-$500 $\mu$m) luminosity (Goudfrooij \& de Jong 1995 and references therein). 
\par\noindent
\footnotemark[$\S$] X-ray luminosity from the $ROSAT$ catalog of O'Sullivan et al. (2001).
}\hss}}
\end{tabular}
\end{center}
\end{table*}
\clearpage

\begin{table*}
\caption{Observation Log}
\begin{center}
\begin{tabular}{lccclcc}
\hline\hline
Name & RA (J2000)&Dec (J2000)&Date &Observing mode &Exposure&Cycles  \\
\hline
NGC~1395&\timeform{3h38m29s.79}&\timeform{-23D01'39''.7}& 2005 Sep 03& MIPS24, large-source photo.&10 sec&2 \\
        &&& 2005 Aug 26& MIPS70, large-source photo.&10 sec&8 \\
        &&& 2005 Sep 03& MIPS160, large-source photo.&10 sec&6 \\
        &&& 2005 Aug 08& IRS SL, staring&60 sec&4\\
NGC~1407 &\timeform{3h40m11s.90}&\timeform{-18D34'49''.4}& 2005 Feb 01& MIPS24, large-source photo.&10 sec&2 \\
        &&& 2005 Aug 27& MIPS70, large-source photo.&10 sec&8 \\
        &&& 2005 Feb 01& MIPS160, large-source photo.&10 sec&6 \\
        &&& 2005 Aug 17& IRS SL, staring&60 sec&4\\
NGC~2974 &\timeform{9h44m33s.28}&\timeform{-3D41'56''.9}& 2005 May 12& MIPS24, large-source photo.&10 sec&2 \\
        &&& 2004 Dec 04& MIPS70, large-source photo.&10 sec&4 \\
        &&& 2005 May 12& MIPS160, large-source photo.&10 sec&6 \\
        &&& 2005 May 22& IRS SL, staring&60 sec&4\\
NGC~3962 &\timeform{11h54m40s.10}&\timeform{-13D58'30''.1}& 2005 Jan 25& MIPS24, large-source photo.&10 sec&2 \\
        &&& 2005 Jan 25& MIPS70, large-source photo.&10 sec&4 \\
        &&& 2005 Jan 25& MIPS160, large-source photo.&10 sec&6 \\
        &&& 2005 Jan 03& IRS SL, staring&60 sec&4\\
IC~3370  &\timeform{12h27m38s.00}&\timeform{-39D20'16''.8}& 2005 Feb 01& MIPS24, large-source photo.&10 sec&2 \\
        &&& 2005 Feb 01& MIPS70, large-source photo.&10 sec&4 \\
        &&& 2005 Feb 01& MIPS160, large-source photo.&10 sec&6 \\
        &&& 2005 Feb 10& IRS SL, staring&60 sec&4\\
NGC~4589 &\timeform{12h37m25s.03}&\timeform{+74D11'30''.8}& 2005 Mar 05& MIPS24, large-source photo.&10 sec&2 \\
        &&& 2004 Nov 06& MIPS70, large-source photo.&10 sec&6 \\
        &&& 2005 Mar 05& MIPS160, large-source photo.&10 sec&6 \\
        &&& 2004 Oct 20& IRS SL, staring&60 sec&4\\
NGC~4696\footnotemark[$*$] &\timeform{12h48m49s.28}&\timeform{-41D18'40''.0}& 2005 Feb 01& MIPS24, small-source photo.&10 sec&1 \\
        &&& 2005 Feb 01& MIPS70, small-source photo.&3 sec&5 \\
        &&& 2005 Feb 01& MIPS160, small-source photo.&3 sec&5 \\
        &&& 2005 Feb 10& IRS SL, staring&60 sec&4\\
\hline
\\
\multicolumn{7}{@{}l@{}}{\hbox to 0pt{\parbox{160mm}{\footnotesize
\par\noindent
\footnotemark[$*$] The MIPS observations of this target were carried out in another GO program.

}\hss}}
\end{tabular}
\end{center}
\end{table*}

\clearpage

\begin{table*}
\caption{Flux Densities of the Observed Galaxies}
\begin{center}
\begin{tabular}{lrrrrrrrrrr}
\hline\hline
Name & \multicolumn{3}{c}{MIPS\footnotemark[$*$]} && \multicolumn{3}{c}{{\it IRAS}\footnotemark[$\dagger$]} && \multicolumn{2}{c}{{\it ISO}\footnotemark[$\ddagger$]}\\
\cline{2-4}\cline{6-8}\cline{10-11}
& 24 $\mu$m & 70 $\mu$m & 160 $\mu$m && 25 $\mu$m & 60 $\mu$m & 100 $\mu$m && 60 $\mu$m & 170 $\mu$m\\
 & (mJy) & (mJy) & (mJy) & & (mJy) & (mJy) & (mJy) & & (mJy) & (mJy)  \\
\hline
NGC~1395 & 39$\pm$5 & 84$\pm$13 & 342$\pm$31 &  & 50$\pm$27 & 100$\pm$25 & 430$\pm$79 & & 80$\pm$30 & 240$\pm$50 \\
NGC~1407 & 40$\pm$5 & 31$\pm$12 & 92$\pm$23 & & 0$\pm$27 & 140$\pm$30 & 430$\pm$65 & & & \\
NGC~2974 & 53$\pm$4 & 678$\pm$19 & 1566$\pm$43 & & 0$\pm$27 & 420$\pm$33 & 1420$\pm$80 & & & \\
NGC~3962 & 19$\pm$4 & 339$\pm$14 & 621$\pm$18 & & 0$\pm$42 & 210$\pm$40 & $<$980 & & 440$\pm$110 & 810$\pm$130 \\
IC~3370 & 35$\pm$6 & 771$\pm$19 & 1150$\pm$27 & & 90$\pm$48 & 570$\pm$28 & 2010$\pm$106 & & & \\
NGC~4589 & 18$\pm$3 & 240$\pm$6 & 394$\pm$9 & & 0$\pm$21 & 210$\pm$31 & 590$\pm$136 & & 210$\pm$100 & 630$\pm$160 \\
NGC~4696 & 24$\pm$4 & 156$\pm$19 & 331$\pm$29 & & 0$\pm$26 & 100$\pm$23 & 740$\pm$131 & & & \\
\hline
\\
\multicolumn{11}{@{}l@{}}{\hbox to 0pt{\parbox{160mm}{\footnotesize
\par\noindent
\footnotemark[$*$] The MIPS flux densities are derived by integrating surface brightness within a diameter of $1.3D_{n}$, where $D_{n}$ is the optical angular diameter listed in Table 1.
\par\noindent
\footnotemark[$\dagger$] The {\it IRAS} flux densities are taken from Knapp et al. (1989), Bregman et al. (1998), or Leeuw et al. (2004).
\par\noindent
\footnotemark[$\ddagger$] The {\it ISO} flux densities are taken from Temi et al. (2004).
}\hss}}
\end{tabular}
\end{center}
\end{table*}

\clearpage

\begin{table*}
\caption{Derived Properties of Dust in the Observed Galaxies}
\begin{center}
\begin{tabular}{lrrrrr}
\hline\hline
Name & \multicolumn{2}{c}{MIPS} && \multicolumn{2}{c}{{\it IRAS}\footnotemark[$*$]}\\
\cline{2-3}\cline{5-6}
& $T_{\rm d}$ & Log$M_{\rm d}$ & &$T_{\rm d}$ & Log$M_{\rm d}$\\
 & (K) & (M$_{\odot}$) & & (K) & (M$_{\odot}$)  \\
\hline
NGC~1395 & 25$\pm$3 & 5.23$\pm$0.20 && 23$\pm$6 & 5.60$\pm$0.03\\
NGC~1407 & 27$\pm$4 & 4.55$\pm$0.27 && $-$ & 5.35$\pm$0.27\\
NGC~2974 & 27$\pm$3 & 6.06$\pm$0.18 && 28$\pm$1 & 6.20$\pm$0.08\\
NGC~3962 & 29$\pm$3 & 5.33$\pm$0.15 && $\ge 32$ & 5.49$\pm$0.08\\
IC~3370 & 31$\pm$3 & 6.02$\pm$0.15 && 29$\pm$1 & 6.73$\pm$0.07\\
NGC~4589 & 30$\pm$3 & 5.20$\pm$0.15 && 31$\pm$4 & 5.62$\pm$0.30\\
NGC~4696 & 27$\pm$3 & 5.62$\pm$0.17 && 24$\pm$3 & 6.67$\pm$0.17\\
\hline
\\
\multicolumn{6}{@{}l@{}}{\hbox to 0pt{\parbox{160mm}{\footnotesize
\par\noindent
\footnotemark[$*$] The {\it IRAS} results are taken from Table 1 of Goudfrooij \& de Jong (1995).
}\hss}}
\end{tabular}
\end{center}
\end{table*}

\clearpage

\begin{table}
\caption{11.3 $\mu$m PAH Emission Features}
\begin{center}
\begin{tabular}{lrcc}
\hline\hline
Name & & Flux (10$^{-21}$ W cm$^{-2}$)\footnotemark[$*$] & EW ($\mu$m)\footnotemark[$\dagger$]\\
\hline
NGC~1395 && 1.86$\pm$0.21 & 0.07$\pm$0.01 \\ 
NGC~1407 && $<$0.3        & $<$0.02        \\
NGC~2974 && 10.48$\pm$0.83 & 0.23$\pm$0.02 \\
NGC~3962 && 5.95$\pm$0.56 & 0.26$\pm$0.02 \\
IC~3370\footnotemark[$\ddagger$]  && 1.27$\pm$0.36 & 0.22$\pm$0.06 \\ 
NGC~4589 && 2.45$\pm$0.33 & 0.14$\pm$0.02 \\
NGC~4696 && $<$0.62       & $<$0.04 \\
\hline
\\
\multicolumn{4}{@{}l@{}}{\hbox to 0pt{\parbox{80mm}{\footnotesize
\par\noindent
\footnotemark[$*$] The uncertainies quoted for the line fluxes are the $1\sigma$ errors from the line fit and do not include the calibration uncertainties. 2 $\sigma$ upper limits are given for non-detected lines.
\par\noindent 
\footnotemark[$\dagger$] EW = equivalent width (observed). 
\par\noindent
\footnotemark[$\ddagger$]] The IRS observation missed the center of the galaxy by $10''$. 
}\hss}}
\end{tabular}
\end{center}
\end{table}

\end{document}